\documentclass[12pt]{article}
\usepackage{graphicx}
\usepackage{epstopdf}
\usepackage{epsfig}
\usepackage{amsmath}
\usepackage{hhline}
\usepackage{amssymb}
\usepackage{times}
\usepackage{cite}
\usepackage{multirow}
\usepackage{color}
\definecolor{rltred}{rgb}{0.75,0,0}
\definecolor{rltgreen}{rgb}{0,0.5,0}
\definecolor{rltblue}{rgb}{0,0,0.75}
\newif\ifpdf
\ifx\pdfoutput\undefined
\else
\fi

\ifpdf
\usepackage{thumbpdf}
\usepackage[pdftex,
        colorlinks=true,
        urlcolor=rltblue,       
        filecolor=rltgreen,     
        linkcolor=rltred,       
        citecolor=rltblue,
        pdftitle={Example File for pdflatex},
        pdfauthor={H1},
        pdfsubject={Template file},
        pdfkeywords={High-Energy Physics, Particle Physics},
        pdfpagemode=None,
        bookmarksopen=true]{hyperref}
 
\fi

\newlength{\dinwidth}
\newlength{\dinmargin}
\begin{document}
%
%
\newcommand{\qsq}{\ensuremath{Q^2} }
\newcommand{\gevsq}{\ensuremath{\mathrm{GeV}^2} }
%
\newcommand{\dstar}{\ensuremath{D^*}}
\newcommand{\dstarp}{\ensuremath{D^{*+}}}
\newcommand{\dstarm}{\ensuremath{D^{*-}}}
\newcommand{\dstarpm}{\ensuremath{D^{*\pm}}}
\newcommand{\zDs}{\ensuremath{z(\dstar )}}
\newcommand{\Wgp}{\ensuremath{W_{\gamma p}}}
\newcommand{\ptds}{\ensuremath{p_t(\dstar )}}
\newcommand{\etads}{\ensuremath{\eta(\dstar )}}
\newcommand{\ptj}{\ensuremath{p_t(\mbox{jet})}}
\newcommand{\ptjn}[1]{\ensuremath{p_t(\mbox{jet$_{#1}$})}}
\newcommand{\etaj}{\ensuremath{\eta(\mbox{jet})}}
\newcommand{\detadsj}{\ensuremath{\eta(\dstar )\, \mbox{-}\, \etaj}}
\def\Journal#1#2#3#4{{#1} {\bf #2} (#3) #4}
\def\NCA{\em Nuovo Cimento}
\def\NIM{\em Nucl. Instrum. Methods}
\def\NIMA{{\em Nucl. Instrum. Methods} {\bf A}}
\def\NPB{{\em Nucl. Phys.}   {\bf B}}
\def\PLB{{\em Phys. Lett.}   {\bf B}}
\def\PRL{\em Phys. Rev. Lett.}
\def\PRD{{\em Phys. Rev.}    {\bf D}}
\def\ZPC{{\em Z. Phys.}      {\bf C}}
\def\EJC{{\em Eur. Phys. J.} {\bf C}}
\def\CPC{\em Comp. Phys. Commun.}

\begin{titlepage}

\noindent
\begin{flushleft}
{\tt DESY 08-080    \hfill    ISSN 0418-9833} \\
{\tt July 2008}                  \\
\end{flushleft}


\vspace{2cm}
\noindent

\begin{center}
\begin{Large}

{\bf Study of Charm Fragmentation into {\boldmath $D^{* {\mathbf 
\pm}}$} Mesons in Deep-Inelastic Scattering at HERA}

\vspace{2cm}

H1 Collaboration

\end{Large}
\end{center}

\vspace{2cm}

\begin{abstract}
\noindent
The process of charm quark fragmentation is studied using $D^{*\pm}$ meson
production in deep-inelastic scattering as measured by the H1 detector at
HERA.  The parameters of fragmentation functions are extracted for QCD
models based on leading order matrix elements and DGLAP or CCFM evolution
of partons together with string fragmentation and particle decays.
Additionally, they are determined for a next-to-leading order QCD
calculation in the fixed flavour number scheme using the independent
fragmentation of charm quarks to $D^{*\pm}$ mesons.
Two different regions of phase space are investigated defined by
the presence or absence of a jet containing the $D^{*\pm}$ meson in the
event. The fragmentation parameters extracted for the two
phase space regions are found to be different.
\end{abstract}

\vspace{1.5cm}

\begin{center}
Submitted to \EJC 
\end{center}

\end{titlepage}

%
%
%
\begin{flushleft}

F.D.~Aaron$^{5,49}$,           
C.~Alexa$^{5}$,                
V.~Andreev$^{25}$,             
B.~Antunovic$^{11}$,           
S.~Aplin$^{11}$,               
A.~Asmone$^{33}$,              
A.~Astvatsatourov$^{4}$,       
A.~Bacchetta$^{11}$,           
S.~Backovic$^{30}$,            
A.~Baghdasaryan$^{38}$
% ...

\bigskip{\it
% ...
$ ^{4}$ Inter-University Institute for High Energies ULB-VUB, Brussels;
          Universiteit Antwerpen, Antwerpen; Belgium$^{ c}$ \\
 $ ^{5}$ National Institute for Physics and Nuclear Engineering (NIPNE) ,
          Bucharest, Romania \\
% ...
$ ^{11}$ DESY, Hamburg, Germany \\
% ...
$ ^{25}$ Lebedev Physical Institute, Moscow, Russia$^{ e}$ \\
% ...
$ ^{30}$ Faculty of Science, University of Montenegro,
          Podgorica, Montenegro$^{ e}$ \\
% ...
$ ^{33}$ Dipartimento di Fisica Universit\`a di Roma Tre
          and INFN Roma~3, Roma, Italy \\
% ...
$ ^{38}$ Yerevan Physics Institute, Yerevan, Armenia \\
% ...
$ ^{49}$ Also at Faculty of Physics, University of Bucharest,
          Bucharest, Romania \\
% ...
$ ^c$ Supported by FNRS-FWO-Vlaanderen, IISN-IIKW and IWT
      and  by Interuniversity
Attraction Poles Programme,
      Belgian Science Policy \\
% ...
$ ^e$ Supported by the Deutsche Forschungsgemeinschaft \\
% ...
}
\end{flushleft}

\newpage
\section{Introduction}
The production of charm quarks is expected to be well described by 
perturbative Quantum Chromodynamics (pQCD) due to the presence of a 
hard scale provided by the charm mass. The evolution of an ``off-shell''
charm quark via gluon radiation until it is ``on-shell'' can be calculated
in pQCD in fixed order of the strong coupling or by summing all orders in
the leading-log approxi\-mation. The transition of an on-shell charm quark
into a charmed hadron is, however, not calculable within the framework of pQCD and is thus usually described by phenomenological models. One of the
major characteristics of this transition process is the longitudinal
momentum fraction transferred from the quark to the hadron, the
distribution of which is parametrised by a fragmentation function. 
\par
Several phenomenological models are available to describe the
transition of a quark into hadrons, for example the independent 
fragmentation~\cite{feynman-field}, the string~\cite{string}, and the
cluster model~\cite{cluster_model}. 
The fragmentation function is unambiguously defined
in a given context of a phenomenological model together with a pQCD
calculation. Universality is then only expected to hold within this context.
\par
The fragmentation function is not a directly measurable quantity as the momentum of the heavy quark is experimentally not accessible. Also the
momentum distribution of the heavy hadron can only be measured within 
a restricted phase space. The momentum spectrum is further affected by
the fact that some heavy hadrons are not produced directly, but are the
result of decays of still heavier excited states, whose contribution is 
not well known. 
\par
The production of charmed hadrons has been measured
and parameters of fragmentation functions have been determined in $e^+e^-$ annihilation experiments~\cite{epem-articles}.
The H1 and ZEUS collab\-orations have published total cross sections for 
the production of various charmed hadrons in deep-inelastic $ep$  scattering (DIS)~\cite{h1-fragfrac} and in 
photoproduction~\cite{zeus-fragfrac}. 
These data show that the probabilities of charm quarks to fragment 
into various final state hadrons are consistent, within experimental 
uncertainties, for $e^+e^-$ and $ep$ collisions.   
\par
In this paper the transition of a charm quark into a $D^{*\pm}$ meson 
in DIS is further investigated. The normalised differential cross 
sections are measured as a function of two observables with different sensitivity to gluon emissions. The momentum of the charm quark is
approximated either by the momentum of the jet, which includes the 
$D^{*\pm}$ meson, or by the sum of the momenta of particles belonging 
to a suitably defined hemisphere containing the $D^{*\pm}$ meson.  
The measurements are performed for two different event samples. The DIS phase space and the kinematic requirements on the $D^{* \pm}$ meson are
the same for both samples. In the first sample, referred to as the
``$D^{* \pm}$ jet sample", the presence of a jet containing the 
$D^{* \pm}$ meson and exceeding a minimal transverse momentum is required
as a hard scale. In the second sample no such jet is allowed to be present. This
sample, referred to as the ``no $D^{* \pm}$ jet sample", allows the investigation
of charm fragmentation in a region close to the kinematic
threshold of charm production.
\par
The normalised differential cross sections are used to fit 
parameters of different fragmentation functions: 
for QCD models as implemented in the 
Monte Carlo (MC) programs RAPGAP~\cite{RAPGAP} and CASCADE~\cite{CASCADE},
which use the Lund string model for fragmentation as implemented in 
PYTHIA~\cite{PYTHIA62}, and 
for a next-to-leading order (NLO) QCD calculation as implemented in HVQDIS~\cite{hvqdis} with the addition of independent fragmentation of
charm quarks to $D^{* \pm}$ mesons.
\par
The paper is organised as follows. Section~\ref{Section:H1Detector} gives
a brief description of the H1 detector. It is followed by the details of
the event selection, the $D^{*\pm}$ meson signal extraction and the jet
selection in section~\ref{Section:DataSelection}. The experimental
fragmentation observables are defined in section~\ref{Section:Observables}.
The QCD models and calculations used for data corrections and for the 
extraction of fragmentation functions are described in 
section~\ref{Section:QCDModels}. The data correction procedure and the 
determination of systematic uncertainties is explained in 
section~\ref{Section:CorrectionsSystematics}. In 
section~\ref{Section:Results} the results of the measurements and of 
the fits of the fragmentation parameters are given. 

\section{H1 Detector}
\label{Section:H1Detector}

The data were collected with the H1 detector at HERA in the years 1999 
and 2000. During this period HERA collided positrons of energy 
$E_{e}=27.5$~GeV with protons of energy $E_{p}=920$~GeV, corresponding to 
a centre-of-mass energy of $\sqrt{s}=319$~GeV. The data sample used for
this analysis corresponds to an integrated luminosity of $47$~pb$^{-1}$.
\par
A right handed Cartesian coordinate system is used with the origin at the nominal primary $ep$ interaction vertex. The direction of the proton beam
defines the positive $z$-axis (forward direction). Transverse momenta are
measured in the $x$-$y$ plane. Polar ($\theta$) and azimuthal ($\phi$)
angles are measured with respect to this reference system. The pseudorapidity is defined as $\eta = - \ln (\tan\frac{\theta}{2})$.
\par
A detailed description of the H1 detector can be found 
in~\cite{H1detector}. Here only the  components relevant for this 
analysis are described. The scattered positron is identified and measured
in the SpaCal~\cite{h1spacal}, a lead-scintillating fibre calorimeter 
situated in the backward region of the H1 detector, covering the
pseudorapidity range $-4.0 < \eta < -1.4$. The SpaCal also provides 
information to trigger on the scattered positron in the kinematic region
of this analysis. Hits in the backward drift chamber (BDC) are used to
improve the identification of the scattered positron and the measurement
of its angle~\cite{h1bdc}. Charged particles emerging from the 
interaction region are measured by the Central Silicon Track detector 
(CST)~\cite{h1cst} and the Central Tracking Detector (CTD), which 
covers a range $-1.74 < \eta < 1.74$. The CTD comprises two large 
cylindrical Central Jet drift Chambers (CJCs) and two $z$-chambers 
situated concentrically around the beam-line, operated within a solenoidal 
magnetic field of $1.16$~T. The CTD also provides triggering information
based on track segments measured in the $r$-$\phi$-plane of the CJCs and 
on the $z$-position of the event vertex obtained from the double layers
of two Multi-Wire Proportional Chambers (MWPCs). The tracking detectors
are surrounded by a finely segmented Liquid Argon calorimeter 
(LAr)~\cite{h1cal}. It consists of an electromagnetic section with lead
absorbers and a hadronic section with steel absorbers and covers the
range $-1.5 < \eta < 3.4$.
\par
The luminosity determination is based on the measurement of the 
Bethe-Heitler process $ep \rightarrow ep\gamma$, where the photon is 
detected in a calorimeter close to the beam pipe at $z = -103$~m.

\section{Data Selection and Analysis}
\label{Section:DataSelection}

The events selected in this analysis are required to contain a scattered
positron in the SpaCal and at least one $D^{*\pm}$ meson candidate. The
scattered positron must have an energy above $8$~GeV. The virtuality of
the photon $Q^{2}$ and the inelasticity $y$ are determined from the
measured energy $E_e^\prime$ and the polar angle $\theta_e^{\prime}$ of
the scattered positron via the relations:

\begin{equation}
Q^2=4E_eE^\prime_e\cos^2\left(
\frac{\theta_e^{\prime}}{2}\right) \; \; \; {\rm and} \; \; \;  
y=1-\frac{ E^\prime_e}{ E_e} \sin^{2} \left(\frac{\theta_e^{\prime}}
  {2}\right)  \; . 
\end{equation}
In addition, the energy $W$ of the $\gamma^{*}p$ rest-frame is determined
using:

\begin{equation}
W^2=ys - Q^2 \; ,
\end{equation}
where $s=4E_eE_p$ is the centre-of-mass energy squared of the $ep$ system.
The photon virtuality is required to be in the range
$2 < Q^{2} < 100$~GeV$^{2}$. This kinematic range is determined by the
geometric acceptance of the SpaCal. The inelasticity is required to lie
in the region $0.05 < y < 0.7$. The difference between the total energy
$E$ and the longitudinal component $P_{z}$ of the total 
momentum, as calculated from the scattered positron and the hadronic final
state, is restricted to $40 < E-P_{z} < 75$~GeV. This requirement
suppresses photoproduction background, where a hadron is misidentified
as the scattered positron. It also reduces the contribution of DIS events
with hard initial state photon radiation, where the positron or photon
escapes in the negative $z$-direction. This leads to values of $E-P_{z}$
significantly lower than the expectation $2E_e = 55$~GeV.
\par
The $D^{*\pm}$ mesons are reconstructed from tracks using the decay channel
$D^{*+} \rightarrow D^{0} \pi_{\rm s}^{+} \rightarrow (K^{-}\pi^{+})\pi_{\rm s}^{+}$ and its charge conjugate, where $\pi_{\rm s}$ denotes the
low momentum pion from the $D^{*\pm}$ meson decay. Requirements on the
transverse momentum and pseudorapidity of the $D^{*\pm}$ meson candidate
and its decay products, as well as on particle identification using 
d$E/$d$x$, are similar to those used in previous H1 
analyses~\cite{h1-dstar}. A summary of the most important requirements
is given in table~\ref{table:dstar-cuts}.

\begin{figure}[hhh]
\center
\epsfig{file=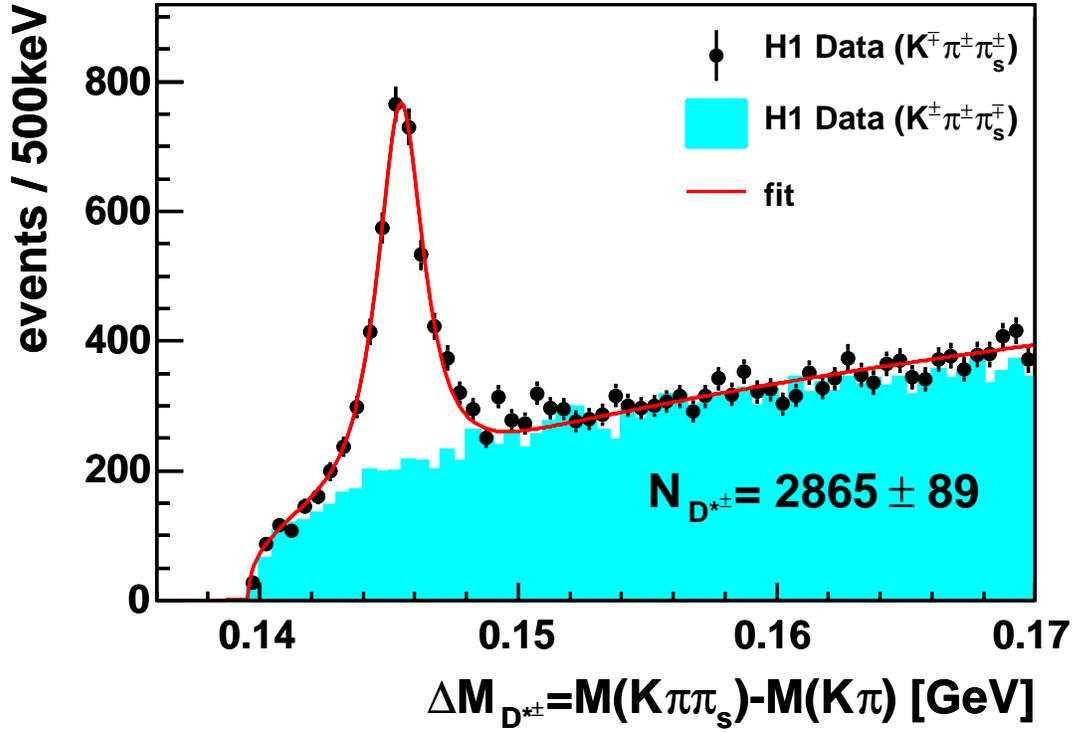,width=\textwidth}
\setlength{\unitlength}{1cm}
\caption{ Distributions of 
$\Delta M_{D^{* \pm}} = M(K\pi\pi_{s}) - M(K\pi)$ for right charge
combinations ($K^{\mp}\pi^{\pm}\pi_{\rm s}^{\pm}$) and for wrong charge
($K^{\pm}\pi^{\pm}\pi_{\rm s}^{\mp}$) combinations in the accepted 
$D^{0}$ mass window. }
\label{fig:deltam} 
\end{figure}

\renewcommand{\arraystretch}{1.15} 
\begin{table}[htdp]
\begin{center}
\begin{tabular}{|c|l|}
\hline
$D^{0}$   &   $P_{\rm T}(K) > 0.25$~GeV \\
          &   $P_{\rm T}(\pi) > 0.25$~GeV \\
                  &   $P_{\rm T}(K) + P_{\rm T}(\pi) > 2$~GeV \\
                  &   $| M(K\pi) - M(D^{0}) | < 0.07$~GeV\\
\hline
$D^{* \pm}$    &   $P_{\rm T}(\pi_{\rm s}) > 0.12$~GeV \\
                  &   $| \eta(D^{*\pm}) |  <  1.5$ \\
                  &   $1.5 < P_{\rm T}(D^{*\pm}) < 15$~GeV \\
\hline
\end{tabular}
\caption{ Kinematic requirements for the selection of $D^{*\pm}$ meson 
candidates. }
\label{table:dstar-cuts}
\end{center}
\end{table}
To select $D^{*\pm}$ meson candidates the invariant mass difference
method~\cite{feldman-deltam} is used. The distribution of 
$\Delta M_{D^{*\pm}} = M(K\pi\pi_{\rm s}) - M(K\pi)$  is shown in 
figure~\ref{fig:deltam} for the full data sample, together with the wrong
charge $K^{\pm}\pi^{\pm}\pi_{\rm s}^{\mp}$ combinations, using
$K^{\pm}\pi^{\pm}$ pairs in the accepted $D^{0}$ mass range. 
Detailed studies show that the wrong
charge $\Delta M_{D^{*\pm}}$ distribution provides a good description of
the right charge $K^{\mp}\pi^{\pm}\pi_{\rm s}^{\pm}$ combinatorial
background. 
\par
The signal is extracted using a simultaneous fit to the 
$\Delta M_{D^{*\pm}}$ distribution of the right and wrong charge
combinations. The signal is fitted using a modified Gaussian 
function~\cite{modgauss} 

\begin{equation}
G_{\rm mod} \propto N_{D^{*\pm}}\, \exp \left[ -0.5\, x^{1+1/(1+0.5\, x)} 
\right] \; , 
\end{equation}
where $x=|\Delta M_{D^{*\pm}} - M_0|/\sigma$. The signal position $M_0$ 
and width $\sigma$ as well as the number of $D^{*\pm}$ mesons 
$N_{D^{*\pm}}$ are free parameters of the fit. The background is 
parametrised as a power function of the form
$N (a+1)\: (\Delta M_{D^{*\pm}}-m_{\pi})^{a}/(M_\textrm{max}-m_{\pi})^{a+1}$,
with the fit boundaries given by the charged pion mass $m_{\pi}$ and $M_\textrm{max}=0.17$~GeV. The two free parameters $a$ and $N$ determine the shape and
normalisation of the background, respectively. The total event sample is
fitted using six free parameters: three for the modified Gaussian, two for
the normalisation of the right and wrong charge $\Delta M_{D^{*\pm}}$
background distributions and one for the background shape, common for the right and
wrong charge combinatorial background. In total $2865 \pm 89$~(stat.) 
$D^{*\pm}$ mesons are obtained. For the differential distributions, the
number of $D^{*\pm}$ mesons in each measurement bin is extracted using the
same procedure, except that the position of the signal peak and its width
are fixed to the values determined from the fit to the total sample.
\par
The hadronic final state is reconstructed in each event using an energy
flow algorithm. The algorithm combines charged particle tracks and calorimetric energy
clusters, taking into account their respective resolution and geometric
overlap, into so called hadronic objects while avoiding double counting of
energy~\cite{hadroo2}. The hadronic objects corresponding to the three
decay tracks forming the $D^{*\pm}$ meson are removed from the event and
replaced by one hadronic object having the four-momentum vector of the
reconstructed $D^{*\pm}$ meson candidate. The energy of the $D^{*\pm}$ meson is calculated using $M(D^{*\pm}) = 2.010$~GeV~\cite{pdg}.
\par
Jets are found in the $\gamma^{*}p$ rest-frame using the inclusive 
$k_{\rm T}$ cluster algorithm~\cite{incl-kt-algo} with the distance
parameter $R=1$ in the $\eta$-$\phi$ plane. In order to combine hadronic
objects into jets, the E-recombination scheme is applied using the
four-momenta of the objects. The jet containing the $D^{*\pm}$ meson
candidate is referred to as the $D^{*\pm}$ jet and is required to have a jet transverse energy $E_{\rm T}^* > 3$~GeV in the $\gamma^{*}p$
rest-frame\footnote{Kinematic variables with the superscript $^*$ refer
to the rest-frame of the virtual photon ($\gamma^{*}$) and proton. In
this frame the photon direction is taken as the direction of the $z$-axis.
The four-vector of the virtual photon used in the boost calculation is
determined from the measurement of the scattered positron.}.
According to MC simulations, the $D^{*\pm}$ jet is found to be well
correlated with the original direction of the charm or anti-charm quark. The distance
in azimuth-pseudorapidity, 
$\Delta r = \sqrt{\Delta\eta^2 + \Delta\phi^2}$, between the charm quark
jet, found using final state partons (``parton level"), and the 
$D^{*\pm}$ jet, found using final state hadrons (``hadron level"), is
below $0.3$ for $90$\% of all events. The correlation between the 
$D^{*\pm}$ jet at hadron level and the $D^{*\pm}$ jet found using charged
particle tracks and calorimetric clusters (``detector level") is even
better, since most of the energy of these jets is reconstructed from
tracks, which are well measured in the tracking system. The number of 
$D^{*\pm}$ mesons is $1508 \pm 68$~(stat.) in the $D^{*\pm}$ jet sample
and $1363 \pm 54$~(stat.) in the no $D^{*\pm}$ jet sample. 

\section{Definition of Experimental Observables}
\label{Section:Observables}

A standard method to study fragmentation is to measure the differential production
cross section of a heavy hadron (H) as a function of a scaled momentum or
energy. In $e^+e^-$ experiments a customary experimental definition of 
the scaled energy is 
${\rm z}_{e^+e^-} = E_{\rm H}/E_{\rm beam}$, where $E_{\rm beam}$ is the
energy of the beams in the centre-of-mass system. In leading order (LO),
i.e. without gluon emissions, the beam energy is equal to the energy of
the charm or anti-charm quark, which are produced in a colour singlet
state. The differential cross section of heavy hadron production as a
function of ${\rm z}_{e^+e^-}$ is directly related to the fragmentation
function.
\par
In the case of $ep$ interactions the situation is more complex. In DIS
the dominant process for $D^{*\pm}$ meson production at HERA is
photon-gluon fusion $\gamma^{*}g \rightarrow c\bar{c}$~\cite{h1-dstar}.
In this case the $c\bar{c}$ pair is produced in a colour octet state. 
The energy of the charm quark pair depends on the energy of the incoming
photon and gluon. Hadrons produced by initial state gluon emissions and
by fragmentation of the proton remnant are also present in the final
state. 
\par
In this analysis charm fragmentation is studied by measuring the
differential cross sections of $D^{*\pm}$ meson production as a function
of two different observables ${\rm z}_{\rm hem}$ and ${\rm z}_{\rm jet}$
defined below, which are sensitive to the fraction of momentum inherited
by the $D^{*\pm}$ meson from the initial charm quark \cite{zuzana}.

\subsection*{The hemisphere method: 
z ${\mathbf =}$ z$_{\rm {\bold{hem}}}$ }

\begin{figure}[hhh]
\center
\epsfig{file=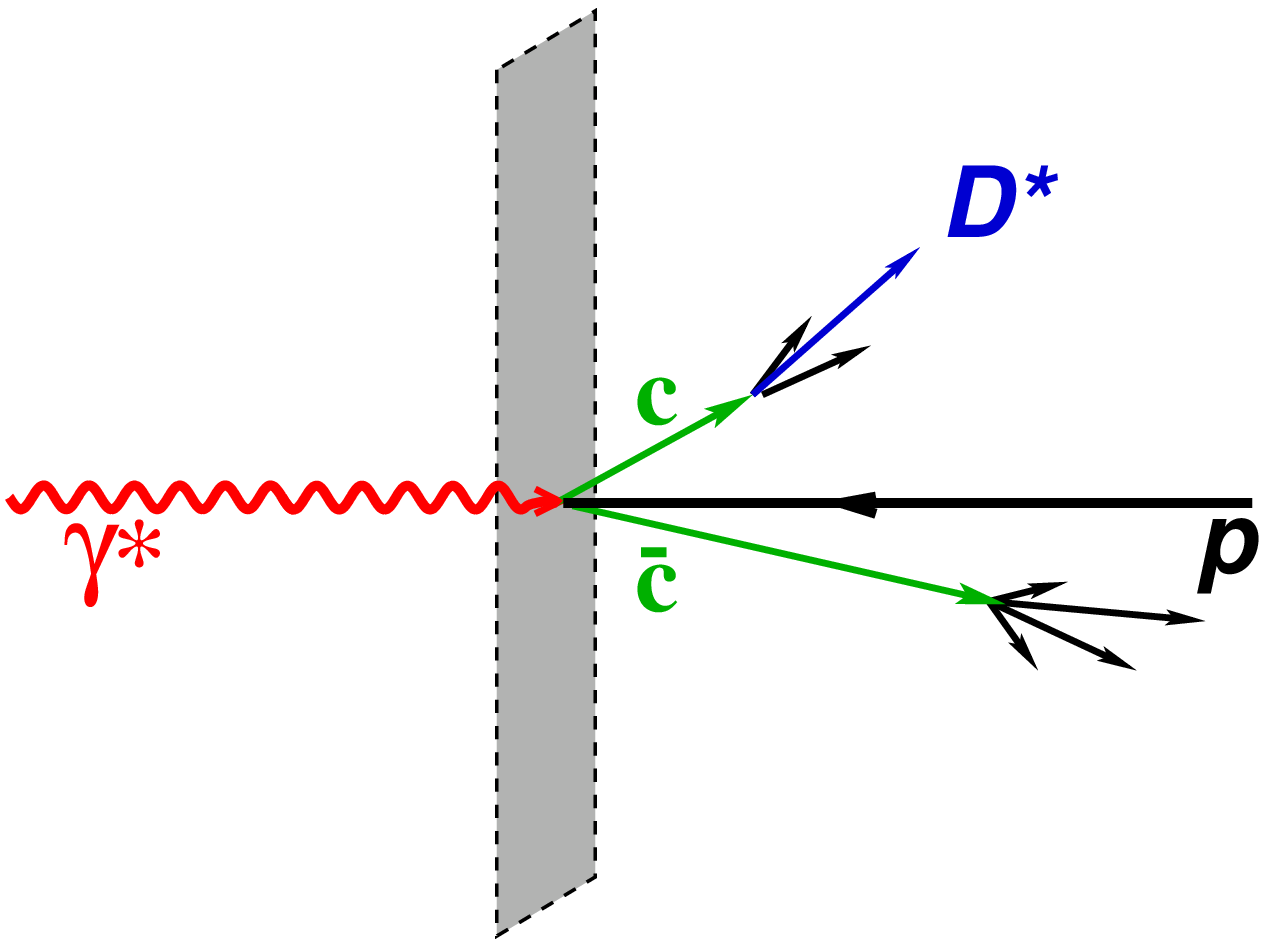,width=0.4\textwidth}
\epsfig{file=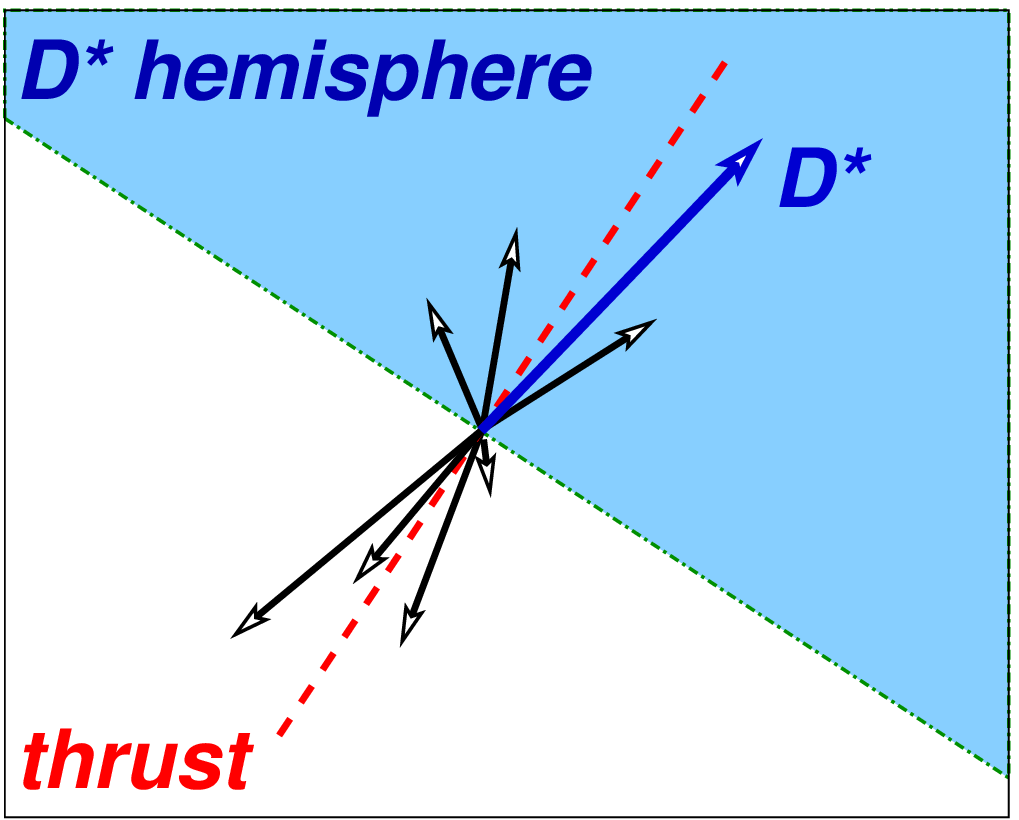,width=0.4\textwidth}
\setlength{\unitlength}{1cm}
\caption{ Illustration of the hemisphere method: a $c{\bar c}$ pair and
the products of its fragmentation in the $\gamma^{*}p$ rest-frame (left)
and in a plane perpendicular to the photon momentum (right). }
\label{fig:hem} 
\end{figure}

In the LO photon-gluon fusion process, which dominates charm
production at HERA, the charm and anti-charm quarks are moving in the
direction of the virtual photon in the $\gamma^*p$ rest-frame of reference. This is due to the fact that the photon is more energetic than the gluon,
which typically carries only a small fraction of the proton's momentum.
Assuming no further gluon radiation in the initial and final state, the
charm and anti-charm quarks are balanced in transverse momentum 
(figure~\ref{fig:hem}, left). This observation suggests to divide the 
event into hemispheres, one containing the fragmentation products of the
charm quark, the other one those of the anti-charm quark. In order to suppress contributions from initial state radiation and the proton
remnant, particles pointing in the proton direction of the $\gamma^{*}p$
rest-frame ($\eta^*<0$) are discarded. The projections of the momenta of
the remaining particles onto a plane perpendicular to the $\gamma^*p$-axis
are determined. Using the projected momenta, the thrust-axis in this plane,
i.e. the axis maximising the sum of the momenta projections onto it, is
found. A plane perpendicular to the thrust-axis allows the division of the
projected event into two hemispheres, one of them containing the $D^{*\pm}$
meson and usually other particles (figure~\ref{fig:hem}, right). The
particles belonging to the same hemisphere as the $D^{*\pm}$ meson are
attributed to the fragmentation of the charm or anti-charm quark. The
fragmentation observable is defined as:

\begin{equation}
 {\rm z}_{\rm hem} = \frac{(E^*+P_{\rm L}^*)_{D^{*\pm}}}{\sum_{\rm hem}
 (E^*+P^*)} \; ,
\label{eq:zhem}
\end{equation}
where in the denominator the energy $E^*$ and the momentum $P^*$ of all
particles of the $D^{*\pm}$ meson hemisphere are summed. The longitudinal
momentum $P^*_{{\rm L}\,D^{*\pm}}$ is defined with respect to the direction
of the three-momentum of the hemisphere, defined as the vectorial sum of
the three-momenta of all particles belonging to the hemisphere. The
variable ${\rm z}_{\rm hem}$ is invariant with respect to boosts along the
direction of the sum of the momenta of all particles in the hemisphere.
Neglecting the mass of the $D^{*\pm}$ meson and of the hemisphere, this definition of ${\rm z}_{\rm hem}$ simplifies to the ratio of their
energies.

\subsection*{ The jet method:  z ${\mathbf =}$ z$_{\rm {\bold{jet}}}$ }

In the case of the jet method the energy and direction of the charm quark
are approximated by the energy and direction of the reconstructed jet,
which contains the $D^{*\pm}$ meson. The fragmentation observable is
defined in analogy to ${\rm z}_{\rm hem}$ as:

\begin{equation}
 {\rm z}_{\rm jet} = \frac{(E^*+P_{\rm L}^*)_{D^{*\pm}}}{(E^*+P^*)_{\rm
  jet}} \; ,
\label{eq:zjet}
\end{equation}
where the longitudinal momentum $P^*_{{\rm L}\,D^{*\pm}}$ is defined with
respect to the direction of the three-momentum of the jet. The jet finding
and the determination of ${\rm z}_{\rm jet}$ are performed in the 
$\gamma^{*}p$ rest-frame.
\par
Both fragmentation observables are defined in such a way that they would
lead to similar distributions, assuming independent fragmentation and no
gluon radiation. The measured distributions, however, are expected to
differ, as they have different sensitivities to gluon radiation and charm
quarks, which are colour connected to the partons of the proton remnant. 
The hemisphere method
typically includes more energy around the charm quark direction than the
jet method. The parameters of fragmentation functions should however be
the same, if extracted for a QCD model, which provides a very good
description of the underlying physics over the full phase space of this
analysis. A comparison of both methods thus may provide a consistency
check and a test of the perturbative and non-perturbative physics as
encoded in the models. 
\par 
The measurement is restricted to  the regions 
$0.2 < {\rm z}_{\rm hem} \leq 1.0$ and 
$0.3 < {\rm z}_{\rm jet} \leq 1.0$, as at lower ${\rm z}$ it is not
possible to separate the $D^{*\pm}$ meson signal from the large
combinatorial background. In order to minimise the sensitivity of the
analysis to the total $D^{*\pm}$ meson cross section, and to reduce
systematic errors, normalised differential cross sections are measured as 
a function of the fragmentation observables ${\rm z}_{\rm hem}$ and 
${\rm z}_{\rm jet}$. The normalisations are chosen such that their
integrals over the respective ${\rm z}$-regions yield unity.

\section{QCD Models and Calculations}
\label{Section:QCDModels}

The MC programs RAPGAP and CASCADE are used to generate events containing
charm and beauty quarks, which are passed through a detailed simulation of
the detector response, based on the GEANT simulation 
program~\cite{Brun:1987ma}. They are reconstructed using the same 
software as used for the data. These event samples are used to 
determine the acceptance and efficiency of the detector and to estimate
the systematic errors associated with the measurements. In addition, these
models are fitted to the data in order to determine the parameters of the
fragmentation functions.
\par
The Monte Carlo program RAPGAP~\cite{RAPGAP}, based on collinear 
factorisation and DGLAP~\cite{dglap} evolution, is used to generate the
direct process of photon-gluon fusion to a heavy $c\bar{c}$~pair, where
the photon acts as a point-like object. In addition, RAPGAP allows the
simulation of charm production via resolved processes, where the photon
fluctuates into partons, one of which interacts with a parton in the
proton, and the remaining partons produce the photon remnant. The program
uses LO matrix elements with massive (massless) charm quarks for the 
direct (resolved) processes. Parton showers based on DGLAP evolution are used to model higher order QCD effects. 
\par
The CASCADE program~\cite{CASCADE} is based on the 
$k_{\rm T}$-factorisation approach. Here, the calculation of the 
photon-gluon fusion matrix element takes into account the charm quark mass
and the virtuality and the transverse momentum of the incoming gluon.
Gluon radiation off the incoming gluon as well as parton showers off the
charm or anti-charm quark are implemented including angular ordering
constraints. The gluon density of the proton is evolved according to the
CCFM equations~\cite{ccfm}. The $k_{\rm T}$-unintegrated gluon density function A0~\cite{CASCADE-gluon}, extracted from inclusive DIS data, is
used. 
\par
In both RAPGAP and CASCADE the hadronisation of partons is performed 
using the Lund string model as implemented in PYTHIA~\cite{PYTHIA62}. 
In the Lund model, the heavy hadron is produced in the process of string
breaking. The fraction of the string longitudinal momentum $z$ carried
by the hadron is generated according to different choices of adjustable 
fragmentation functions  $D^{\rm H}_{ \rm Q}(z)$. Within this analysis
three widely used parametrisations are employed, of which two depend on
a single free parameter, and one depends on two free parameters. The 
parametrisation suggested by Peterson et al.~\cite{peterson-ff} has the
functional form:

\begin{equation}
D^{\rm H}_{ \rm Q}(z) \varpropto \frac{1}{{z}[1-(1/{z})-\varepsilon / 
(1-{z})]^2} \; ,
\label{eq:peterson}
\end{equation}
and the one by Kartvelishvili et al.~\cite{kartvelishvili-ff} is given 
by:

\begin{equation}
D^{\rm H}_{ \rm Q}(z) \varpropto  {z}^{\alpha} (1-{z}) \; .
\label{eq:kartvelishvili}
\end{equation}
The free parameters $\varepsilon$ and $\alpha$ determine the ``hardness"
of the fragmentation function and are specific to the flavour of the heavy
quark, i.e. charm in the case of $D^{*\pm}$ meson production. The
parametrisation inspired by Bowler and Morris~\cite{bowler-ff} (referred
to as the Bowler parametrisation) has the functional form:

\begin{equation}
D^{\rm H}_{ \rm Q}(z) \varpropto \frac{1}{z^{1+r_{\rm Q}bm_{\rm Q}^2}}
(1-z)^a \exp{(- \frac{bM_{\rm T}^2}{z})} \; .
\label{eq:bowler}
\end{equation}
The shape of the fragmentation function is determined by two free 
parameters $a$ and $b$, $m_{\rm Q}$ is the mass of the heavy quark, 
$M_{\rm T}= \sqrt{M^{2}_{\rm H}+P_{\rm T}^{2}}$ the transverse mass of
the heavy hadron, and $r_{\rm Q}=1$ as default in PYTHIA. 
\par
For data corrections the parameter setting tuned by the ALEPH 
collaboration\cite{ALEPH-steering} together with the Peterson fragmentation
function is used for the fragmentation of 
partons in PYTHIA. It includes higher excited charm states, of which some
also decay to $D^{*\pm}$ mesons and contribute significantly to the 
$D^{*\pm}$ meson yield. When extracting parameters of the fragmentation
functions also the default parameter setting of PYTHIA is used as an
alternative. In this case no higher excited charm states are produced. 
Both parameter settings are indicated in table~\ref{table:steering}.
\par
The parameters for the Kartvelishvili and Peterson fragmentation functions
are also extracted for the HVQDIS program~\cite{hvqdis}. HVQDIS is based on 
the NLO, i.e. $\mathcal{O}(\alpha_{\rm s}^2)$, calculation in the fixed
flavour number scheme, with three light active flavours as well as gluons
in the proton. The proton parton density functions (PDFs) of the light 
quarks and the gluon are evolved according to the DGLAP equations. 
Massive charm quarks are assumed to be produced only 
perturbatively via photon-gluon fusion and higher order processes. The
final state charm quarks are fragmented independently into $D^{*\pm}$ mesons in the $\gamma^* p$ rest-frame. Kartvelishvili and Peterson
parametrisations are used to generate the charm quark's momentum fraction
transferred to the $D^{*\pm}$ meson. The energy of the charm quark is calculated using the on-mass-shell condition. In addition, the $D^{*\pm}$
meson can be given a transverse momentum $P_{\rm T}$ with respect to the
charm quark, according to the function $P_{\rm T} \exp(-\beta P_{\rm T})$.
The value used for the parameter $\beta$ corresponds to an average
$P_{\rm T}(D^{*\pm})$ of $350$~MeV.
\par
The Monte Carlo programs RAPGAP and HERWIG~\cite{HERWIG} are used to 
estimate the size of the hadronisation corrections to the data for
comparison with HVQDIS predictions. While the perturbative QCD model of
HERWIG is similar to the one of RAPGAP, the HERWIG program employs the
cluster hadronisation model, which is quite different from the Lund string
model used by PYTHIA. 
\par
The basic parameter choices for the QCD models and the NLO calculation 
are summarised in table~\ref{table:models}.

\section{Data Corrections and Systematic Errors}
\label{Section:CorrectionsSystematics}

\begin{figure}[hhh]
\center{ \Large  \bf {\boldmath $D^{*\pm}$} jet sample}
\center
\vskip -0.4cm
\epsfig{file=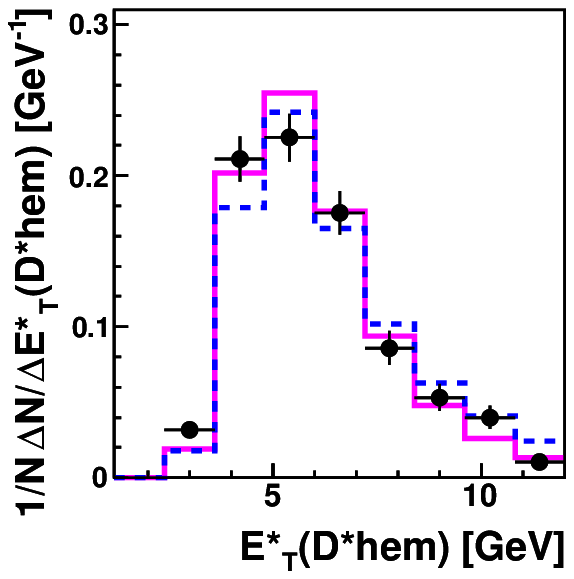,width=0.4\textwidth}
\epsfig{file=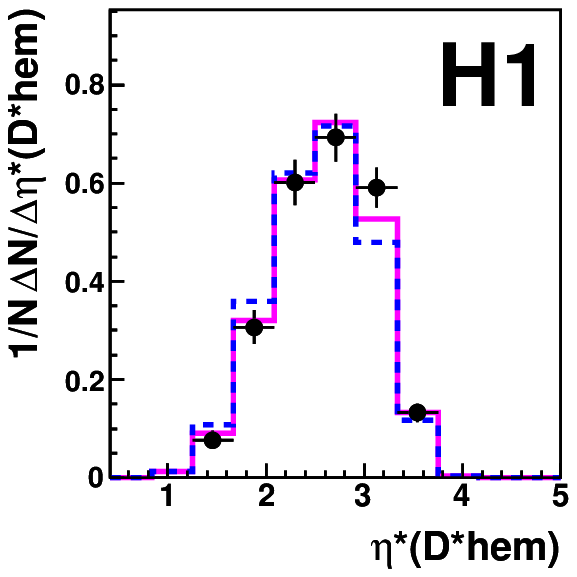,width=0.4\textwidth}
\epsfig{file=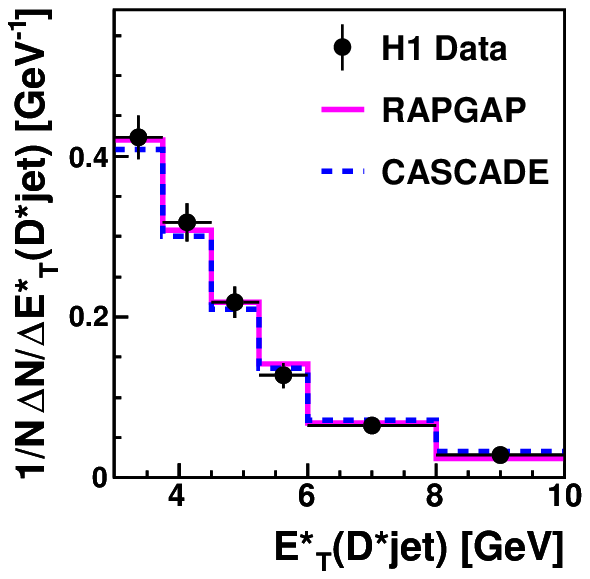,width=0.4\textwidth}
\epsfig{file=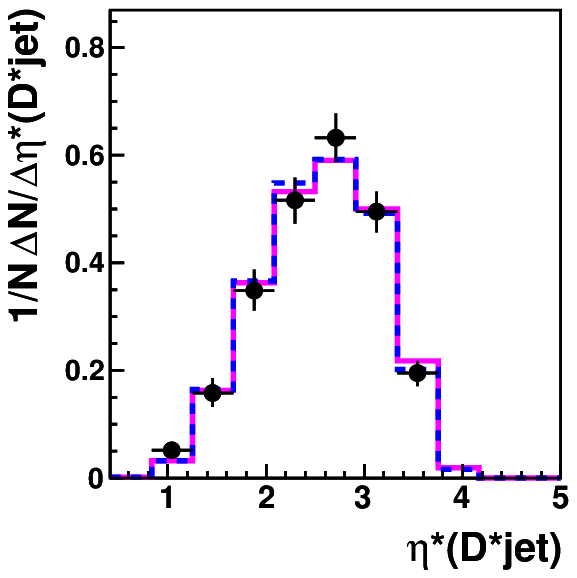,width=0.4\textwidth}
\setlength{\unitlength}{1cm}
\caption{Comparison on detector level between the $D^{*\pm}$ jet data
sample and the reweighted Monte Carlo models (see 
section~\ref{Section:CorrectionsSystematics}) used to correct the data
for detector effects. Shown are $E_{\rm T}^*$ and $\eta^*$ of the 
$D^{*\pm}$ meson hemisphere, calculated from the sum of momenta of all
particles in the hemisphere, and $E_{\rm T}^*$ and $\eta^*$ of the 
$D^{*\pm}$ jet. All observables are calculated in the $\gamma^*p$ 
rest-frame. }
\label{fig:controlplots} 
\end{figure}

\begin{figure}[hhh]
\center{ \Large  \bf No {\boldmath $D^{*\pm}$} jet sample}
\center
\vskip -0.4cm
\epsfig{file=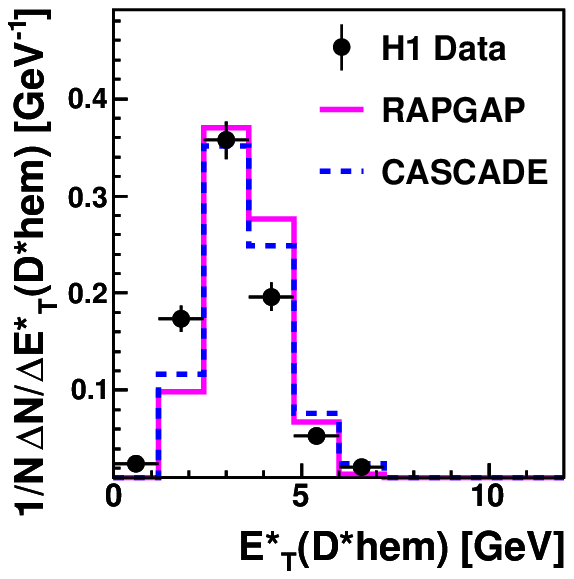,width=0.4\textwidth}
\epsfig{file=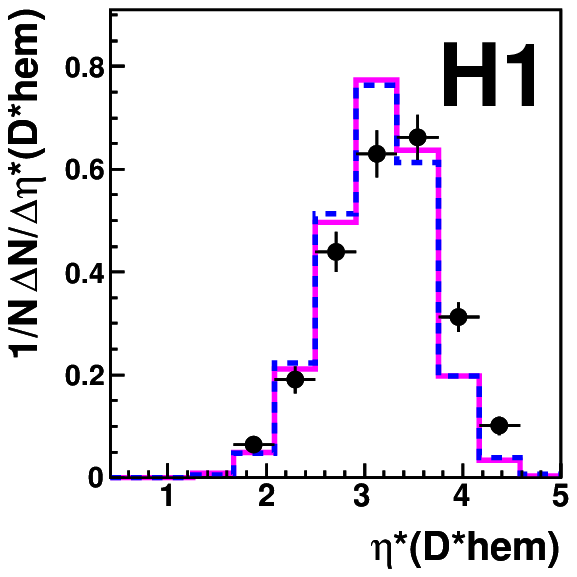,width=0.4\textwidth}
\setlength{\unitlength}{1cm}
\caption{Comparison on detector level between the no $D^{*\pm}$ jet 
sample and reweighted Monte Carlo models (see 
section~\ref{Section:CorrectionsSystematics}) used to correct the data
for detector effects for the no $D^{*\pm}$ jet sample. Shown are
$E_{\rm T}^*$ and $\eta^*$ of the $D^{*\pm}$ meson hemisphere, 
calculated from the sum of momenta of all particles in the hemisphere.
All quantities are calculated in the $\gamma^* p$ rest-frame. }
\label{fig:controlplots-noDsJetsample} 
\end{figure}

In this analysis, the differential cross section for the production of
$D^{*\pm}$ mesons, which result from the fragmentation of charm quarks
either directly or via decays from higher excited charm states, is
measured. The small contribution of $D^{*\pm}$ mesons originating from
B-hadron decays is estimated with RAPGAP and is subtracted from the data. It is on the level of $1$ to $2$\%. The data are corrected for detector
and QED radiative effects. The transverse momentum and pseudorapidity
distributions of the $D^{*\pm}$ mesons of the Monte Carlo event samples,
which are used to correct the data samples for detector effects, are 
reweighted to the corresponding distributions of the data in order to
achieve an improved description. The $\eta$ and $P_{\rm T}$ dependent
reweighting
factors differ from unity by typically $10-30$\%. After this
reweighting, both RAPGAP and CASCADE provide a good description of
the data as shown in figure~\ref{fig:controlplots}. 
The description of the no $D^{*\pm}$ event sample by the reweighted MC models,
as shown in figure~\ref{fig:controlplots-noDsJetsample}, is worse.

The measurement bins are defined
in such a way that the purity in each bin, defined as the fraction of
events reconstructed in a ${\rm z}_{\rm hem}$ or ${\rm z}_{\rm jet}$ bin
that originate from that bin on hadron level, is between $40$ and $70$\%.
\par
The correction for the detector effects is done using regularised
deconvolution, taking into account migrations between measurement
bins~\cite{deconvolution}. The detector response matrix is generated using
RAPGAP, and the value of the regularisation parameter is determined through
decomposition of the data into eigenvectors of the detector response
matrix. As a check, the detector response matrix was also generated using
CASCADE and found to be consistent with the one from RAPGAP. Statistical
errors are calculated by error propagation using the covariance matrix.
The data are then corrected for migrations into the visible phase space
using  RAPGAP and CASCADE. The effects of QED radiation are corrected for
using the HERACLES~\cite{heracles} program, which is interfaced with 
RAPGAP. Correction factors are calculated from the ratio between cross
sections obtained from the model including and not including QED 
radiation. The corrections are applied bin-by-bin in ${\rm z}_{\rm hem}$
and ${\rm z}_{\rm jet}$. In the case of  ${\rm z}_{\rm jet}$ the
corrections are about $2$\%. In the case of ${\rm z}_{\rm hem}$, 
where photons radiated into the $D^{*\pm}$ meson hemisphere can be mistaken 
as fragmentation products of the charm quark, the corrections reach 
$10$\% for the lowest value of ${\rm z}_{\rm hem}$.
\par
In contrast to the QCD models discussed so far, the HVQDIS program 
provides only a partonic final state with the exception of the additional
$D^{*\pm}$ meson from charm fragmentation. In order to compare the NLO
predictions to the measurements, the data are corrected to the parton
level by means of hadronisation corrections, which are estimated using
the MC generators RAPGAP and HERWIG. While the quantity
$(E^*+P_{\rm L}^*)_{D^{*\pm}}$ in equation~\ref{eq:zhem} and~\ref{eq:zjet}
is calculated using the momentum of the $D^{*\pm}$ meson, the jet finding
and the calculation of the jet and hemisphere quantities, the denominators
in equations~\ref{eq:zhem} and~\ref{eq:zjet}, are performed using the 
partonic final state. All partons after parton
showering are considered, and the same jet and hemisphere finding
algorithms are applied at parton and hadron level. For each $\rm z$-bin
the hadronisation correction factor is calculated as the ratio of parton
to hadron level cross section. The arithmetic mean of the hadronisation
correction factors of both models is used to multiply the data cross
section. In case of $\rm z_{hem}$ the hadronisation corrections differ
from unity by typically $\pm 40$\%. For $\rm z_{jet}$ they differ from
unity by typically $\pm 20$\%, except for the highest $\rm z$-bin, where
they are about $50$\%. The hadronisation corrections as determined by HERWIG and RAPGAP are similar for most of the measurement bins, with 
the exception of the lowest bin in $\rm z_{jet}$, where they differ by 
about $60 $\%.
\par
The following systematic uncertainties on the normalised differential
cross sections are considered:
\begin{itemize}
\item 
The energy uncertainty of the scattered positron varies linearly from 
$\pm 3$\% for an energy of $8$~GeV to $\pm 1$\% for $27$~GeV.
\item 
The polar angle of the scattered positron has an estimated uncertainty 
of $\pm 1$~mrad.
\item 
The uncertainty of the energy scale of the hadronic objects is made up 
of $\pm 0.5$\% due to tracks and $\pm 4$\% ($\pm 7$\%) due to LAr
(Spacal) clusters.
\item 
The effect of the uncertainty of the tracking efficiency on reconstructing 
the $D^{*\pm}$ meson is determined by changing the nominal efficiency
in the simulation as a function of track $\eta$ and $P_{\rm T}$. In the
central region of the accepted $\eta$--$P_{\rm T}$ phase space the
estimated uncertainty of the nominal efficiency is $\pm 2$\%, in the
regions of large $|\eta|$ but not small $P_{\rm T}$ it is $\pm 3$\%,
and for large $|\eta|$ and small $P_{\rm T}$ it is $\pm 4$\%.
\item 
The value of d$E/$d$x$ of the $D^{*\pm}$ meson decay products has an
estimated uncertainty of $\pm 8$\%, which is of similar size as the 
experimental resolution in d$E/$d$x$.
\item 
The uncertainty of the $D^{*\pm}$ meson signal extraction is estimated
using different $D^{*\pm}$ counting techniques and by using different 
fit functions for the background parametrisation.
The largest uncertainty comes from the background description, which
determines the systematic error on the signal extraction.
\item 
The uncertainty of beauty production by the RAPGAP MC is assumed to be 
$\pm 100$\%. The resulting small uncertainty of the normalised $D^{*\pm}$
meson cross sections is taken to be symmetrical.
\item
The effect of using different MC models for the small correction for
migrations into the visible phase space is studied using RAPGAP and
CASCADE. 
The factors used to correct the data are determined as the average of the
correction factors obtained from the two models.
Half the difference is taken 
as systematic uncertainty.
\item
For parton level corrected distributions half the difference between
the hadronisation correction factors of RAPGAP and HERWIG is taken as
the uncertainty due to the different fragmentation models.
\end{itemize}
\par
Other systematic effects, which are investigated and found to be negligible,
are: the effect of reflections, i.e. wrongly or incompletely reconstructed
$D^{*\pm}$ meson decays, on the shape of the fragmentation observables,
the effect on acceptance and reconstruction efficiency from including
diffractive events, the effect of using different MC models for the
deconvolution of the data and the uncertainty of the QED radiative
effects.
\par
Each source of systematic error is varied in the Monte Carlo simulation
within its uncertainty. In each measurement bin, the corresponding
deviation of the normalised cross sections from the central value is taken
as the systematic error. Among the systematic errors the uncertainties due to
the scattered positron energy scale, the hadronic energy scale, and the
beauty fraction are correlated amongst the bins in ${\rm z}$. In the
extraction of the parameters of the fragmentation functions, the
statistical and systematic errors with their correlations are taken into
account. The average effect of various systematic errors on the 
${\rm z}_{\rm hem}$ and ${\rm z}_{\rm jet}$ distributions is summarised
in table~\ref{table:systematics}. Since the distributions of
${\rm z}_{\rm hem}$ and ${\rm z}_{\rm jet}$ are normalised, the effect of
many systematic uncertainties is reduced and the statistical error
dominates the uncertainty of the measurement.

\section{Results}
\label{Section:Results}

\subsection{Normalised differential cross sections and comparison with 
predictions}
\label{Subsection:results_default}

\begin{figure}[hhh]
\center
\epsfig{file=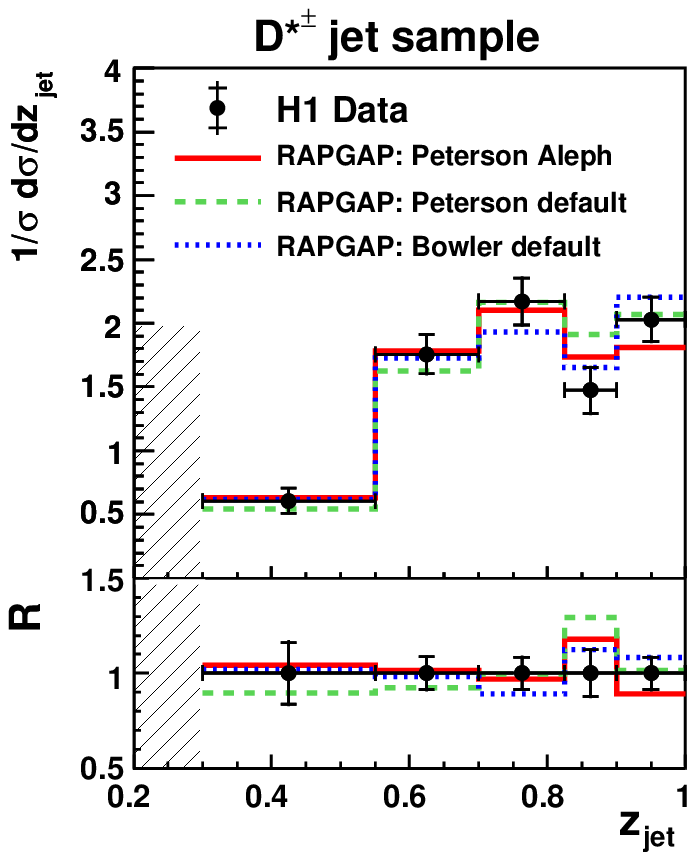,width=0.49\textwidth}
\epsfig{file=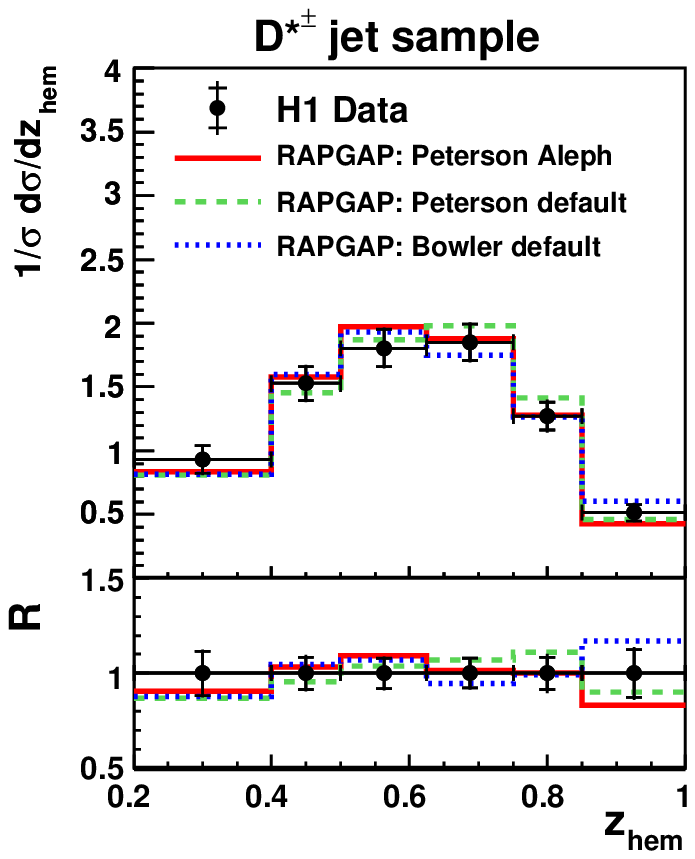,width=0.49\textwidth}
\setlength{\unitlength}{1cm}
\caption{ Normalised $D^{*\pm}$ meson cross sections as a function of 
${\rm z}_{\rm jet}$ and ${\rm z}_{\rm hem}$ for the $D^{*\pm}$ jet 
sample. The measurements are normalised to unity in the displayed range
of ${\rm z}_{\rm jet}$ and ${\rm z}_{\rm hem}$, respectively. The data
are compared with MC predictions of RAPGAP, using PYTHIA default settings
with Peterson or Bowler parametrisations and the ALEPH setting, which
includes the production of higher excited charm states (see 
table~\ref{table:steering}). The ratio ${\rm R= MC/data}$ is shown as
well as the relative statistical uncertainties (inner error bars) and
the relative statistical and systematic uncertainties added in quadrature (outer error bars) for the data points put to ${\rm R}=1$. }
\label{fig:results-defaults} 
\end{figure}

The differential cross sections of $D^{*\pm}$ meson production as a
function of the fragmentation observables ${\rm z}_{\rm hem}$ and 
${\rm z}_{\rm jet}$ are shown in figure~\ref{fig:results-defaults} for 
the $D^{*\pm}$ jet sample. They refer to the visible phase space given 
by $2 < Q^2 < 100$~GeV$^2$, $0.05 < y < 0.7$,
$1.5 < P_{\rm T}(D^{*\pm}) < 15$~GeV and $ |\eta(D^{*\pm}) | < 1.5$. In
addition, a $D^{*\pm}$ jet with $E_{\rm T}^* > 3$~GeV in the $\gamma^{*}p$
rest-frame is required in order to have the same hard scale in the event for both distributions ${\rm z}_{\rm hem}$ and ${\rm z}_{\rm jet}$. The
measurements and the corresponding predictions are normalised such that
their integrals over the respective  ${\rm z}$-regions yield unity. 
The normalised cross sections and
their errors are given in table~\ref{table:results-hem-dsjet} for the
hemisphere observable and in table~\ref{table:results-jet} for the jet observable. 
\par
Figure~\ref{fig:results-defaults} also includes predictions of RAPGAP with
three commonly used fragmentation parameter settings for PYTHIA 
(described in table~\ref{table:steering}), obtained from $e^{+}e^{-}$ annihilation. The
settings and the corresponding values of  $\chi^2$/n.d.f., as calculated
from the data and the model predictions, are summarised in 
table~\ref{table:results-parameters-default}. In general, there is 
reasonable agreement between data and the QCD model with all settings 
for both the jet and the hemisphere observables. 
The large difference between
the two distributions observed in the highest ${\rm z}_{\rm jet}$ bin is
mainly due to a significant fraction of $D^{*\pm}$ jets consisting of a
$D^{*\pm}$ meson only, for which ${\rm z}_{\rm jet}$ equals unity.
CASCADE provides a
similar description of the data as RAPGAP.

\subsection{Extraction of parameters for the Kartvelishvili and 
Peterson fragmentation functions}
\label{subsection:fit}

The normalised $D^{*\pm}$ meson differential cross sections as a 
function of ${\rm z}_{\rm hem}$ and ${\rm z}_{\rm jet}$ are used to 
extract optimal parameters for the Peterson and Kartvelishvili 
fragmentation functions described in section~\ref{Section:QCDModels}.
Both parametrisations have a single free parameter.
\par
\begin{figure}[hhh]
\center
\epsfig{file=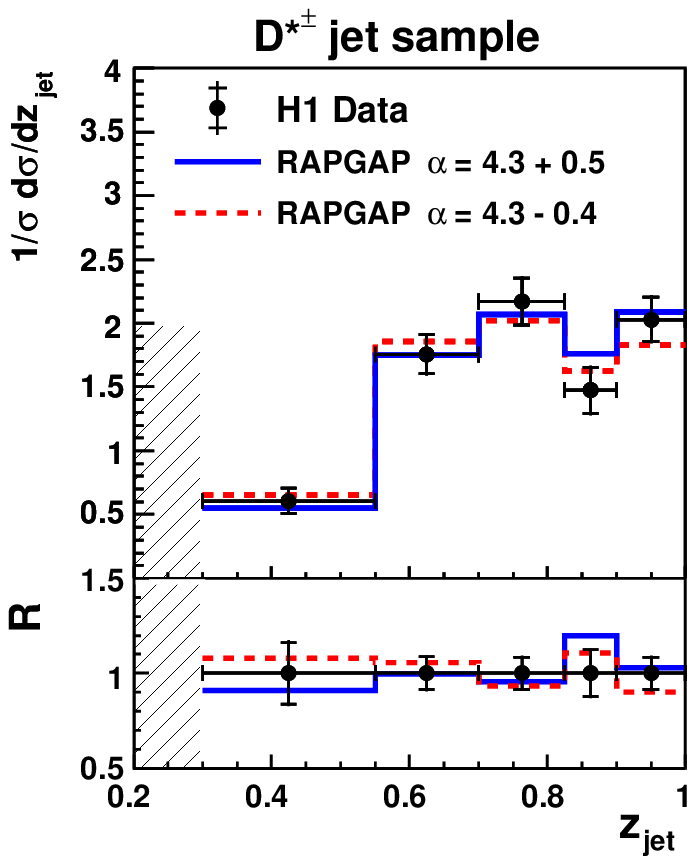,width=0.49\textwidth}
\epsfig{file=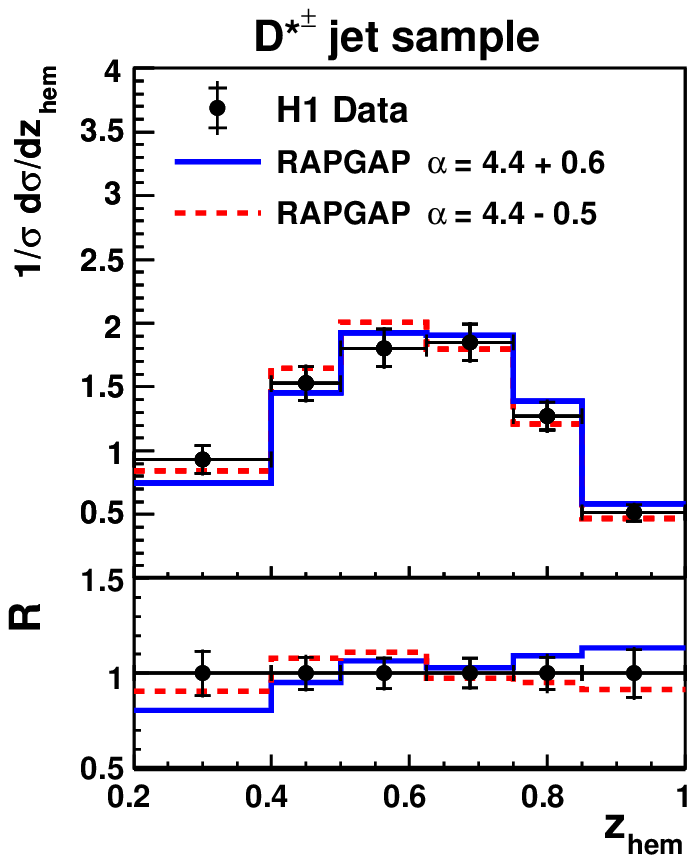,width=0.49\textwidth}
\setlength{\unitlength}{1cm}
\caption{ Normalised $D^{*\pm}$ meson cross sections as a function of 
${\rm z}_{\rm jet}$ and ${\rm z}_{\rm hem}$ for the $D^{*\pm}$ jet sample.
The measurements are normalised to unity in the displayed range of 
${\rm z}_{\rm jet}$ and ${\rm z}_{\rm hem}$, respectively. The same data as in figure~\ref{fig:results-defaults} are compared to the predictions of
the MC program RAPGAP with the ALEPH setting for PYTHIA and Kartvelishvili
parametrisation using the fragmentation function parameter $\alpha$ fitted
according to the procedure described in section~\ref{Section:Results}. The
full and dashed lines indicate a variation of the fragmentation parameter
by $\pm 1 \sigma$ around the best fit value of $\alpha$. The ratio 
${\rm R=  MC/data}$ is described in the caption of 
figure~\ref{fig:results-defaults}.}
\label{fig:results-rapgap-kart} 
\end{figure}

The parameter extraction is done by comparing different model 
configurations with the data. A configuration is defined by one of the QCD 
calculations (RAPGAP, CASCADE or HVQDIS), by one of the fragmentation 
functions (Peterson or Kartvelishvili) and by a possible value for the 
corresponding fragmentation parameter, $\varepsilon$ or $\alpha$. For
RAPGAP and CASCADE the configuration also depends on the PYTHIA parameter settings used (ALEPH and default, see table~\ref{table:steering}).
In order to be able to compare all configurations to the data, a reweighting 
procedure is applied. For each of the QCD calculations large event samples
with $D^{*\pm}$ mesons are generated using the Peterson fragmentation
function. For these events the $z$-value of the fragmentation function,
used by the model to generate the fraction of charm quark or string
momentum transferred to the $D^{*\pm}$ meson, is stored such that each
event can be reweighted to another fragmentation function or any other
parameter value. For each configuration the predicted and measured distributions of the fragmentation observables are used to determine a 
$\chi^2$ as a function of the fragmentation parameter. In the calculation
of the $\chi^2$ the full covariance matrix is used, taking into account
correlated and uncorrelated statistical and systematic errors. 
The fragmentation parameter is determined at the minimum of the $\chi^2$.
The shape of the $\chi^2$ distribution is used to determine the 
$\pm 1\sigma$ error (using $\chi^2_{\rm min} + 1$) of the extracted parameter. As an example, in figure~\ref{fig:results-rapgap-kart} the
data are compared to the prediction of RAPGAP with the ALEPH setting for
PYTHIA as given in table~\ref{table:steering} but using the Kartvelishvili
parametrisation. The two lines indicate the $\pm 1 \sigma$ total
uncertainty around the best fit value of $\alpha$. The description of 
the data by CASCADE is similar.
\par
The parameters $\alpha$ and $\varepsilon$, which are extracted using
RAPGAP and CASCADE, with and without higher excited charmed hadrons, 
are summarised in table~\ref{table:results-parameters-hem-jet} together 
with their corresponding values of $\chi^2$/n.d.f.. With the fitted 
parameters the model predictions using either the Peterson or the
Kartvelishvili parametrisations describe the data reasonably well, with 
the Kartvelishvili parametrisation being in all cases slightly preferable,
as indicated by the values of $\chi^2$/n.d.f.. When using the same PYTHIA
parameter setting, the fragmentation parameters $\alpha$ and 
$\varepsilon$, extracted from the ${\rm z}_{\rm hem}$ and 
${\rm z}_{\rm jet}$ observables, are in good agreement. Both RAPGAP and
CASCADE lead to statistically compatible parameters $\varepsilon$ and 
$\alpha$. A priori, agreement in the fragmentation function parameters 
for RAPGAP and CASCADE is not required, since the models differ in terms
of simulated processes (direct and resolved in case of RAPGAP compared to
direct only for CASCADE) and in their implementation of perturbative QCD. 
\par
The fragmentation parameters $\alpha$ and $\varepsilon$ depend
significantly on the PYTHIA parameter settings used, i.e. whether 
$D^{*\pm}$ mesons are assumed to be produced only via direct fragmentation
of charm quarks or additionally originate from decays of higher excited
charm states. In the latter case the $D^{*\pm}$ mesons carry a smaller
fraction of the original charm or anti-charm quark momentum in comparison
with the directly produced ones. Both the default PYTHIA setting and the
setting containing higher excited charm states describe the data equally
well. The values of the Peterson parameter $\varepsilon$ extracted for the 
PYTHIA setting containing higher charm states, see table~\ref{table:results-parameters-hem-jet}, are in agreement with the
value $\varepsilon = 0.04$ tuned by ALEPH\cite{ALEPH-steering}. 
This result is consistent with
the hypothesis of fragmentation universality in $ep$ and $e^+e^-$
processes. 
\par
\begin{figure}[hhh]
\center
\epsfig{file=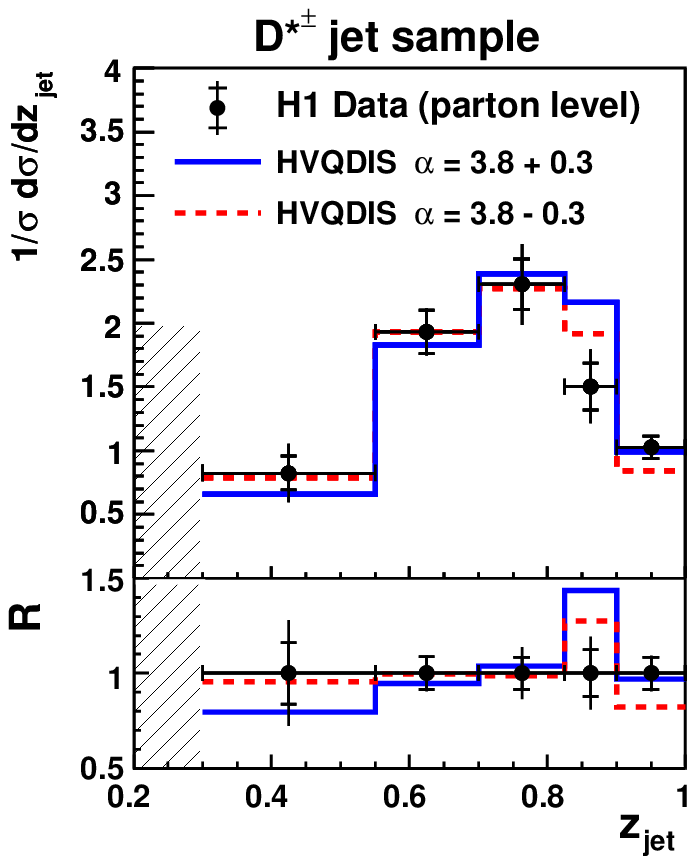,width=0.49\textwidth}
\epsfig{file=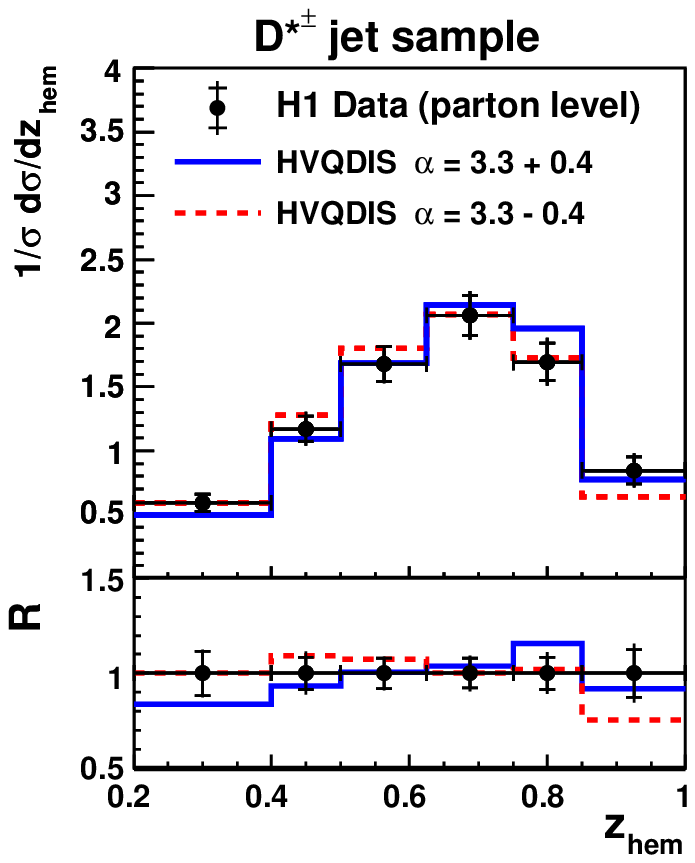,width=0.49\textwidth}
\setlength{\unitlength}{1cm}
\caption{ Normalised $D^{*\pm}$ meson cross sections as a function of 
${\rm z}_{\rm jet}$ and ${\rm z}_{\rm hem}$ for the $D^{*\pm}$ jet sample.
The data are corrected for hadronisation effects (see 
section~\ref{Section:QCDModels}). The measurements are normalised to unity in the displayed range of ${\rm z}_{\rm jet}$ and ${\rm z}_{\rm hem}$,
respectively. The data are compared to NLO predictions of HVQDIS with the
Kartvelishvili parametrisation using the fragmentation function
parameter $\alpha$ fitted according to the procedure described in 
section~\ref{Section:Results}. The full and dashed lines indicate a
variation of the fragmentation parameter by $\pm 1 \sigma$ around the 
best fit value of $\alpha$. The ratio ${\rm R=  MC/data}$ is described in
the caption of figure~\ref{fig:results-defaults}.}
\label{fig:results-hvqdis-kart} 
\end{figure}

The NLO calculation as implemented in HVQDIS with the Kartvelishvili 
fragmentation function leads to a good fit of the data, corrected for
hadronisation effects, as shown in figure~\ref{fig:results-hvqdis-kart}.
On the other hand HVQDIS provides a rather poor description of 
${\rm z}_{\rm hem}$ and ${\rm z}_{\rm jet}$, if the Peterson fragmentation
function is used (the $\chi^2$ values of the fit are shown in 
table~\ref{table:results-parameters-hem-jet}). Simulating a
$P_{\rm T}$ of the $D^{*\pm}$ meson with respect to the charm quark
direction, as explained in section~\ref{Section:QCDModels}, has only a little
effect on the extracted value of $\alpha$.
\par
\begin{figure}[hhh]
\center
\epsfig{file=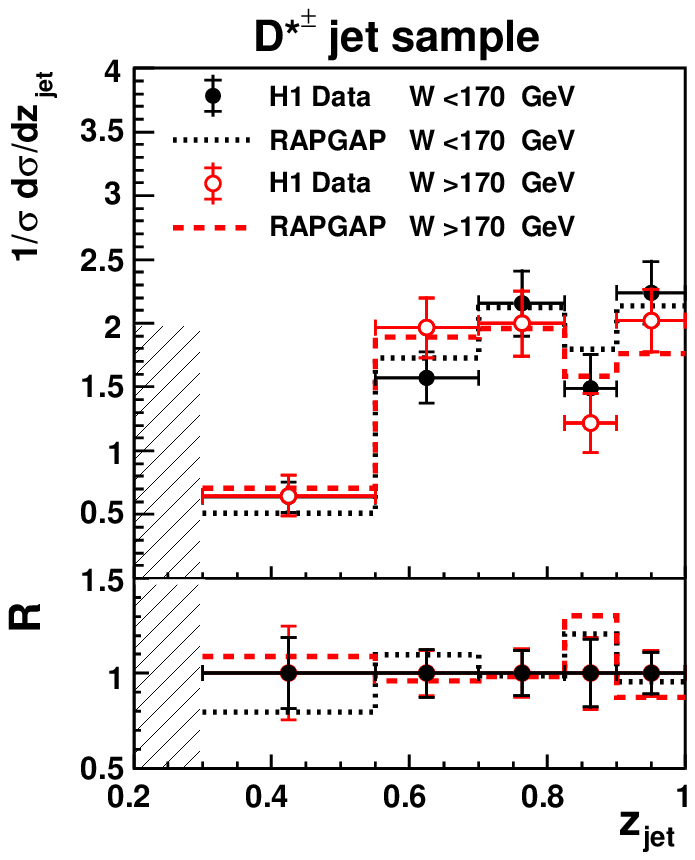,width=0.49\textwidth}
\epsfig{file=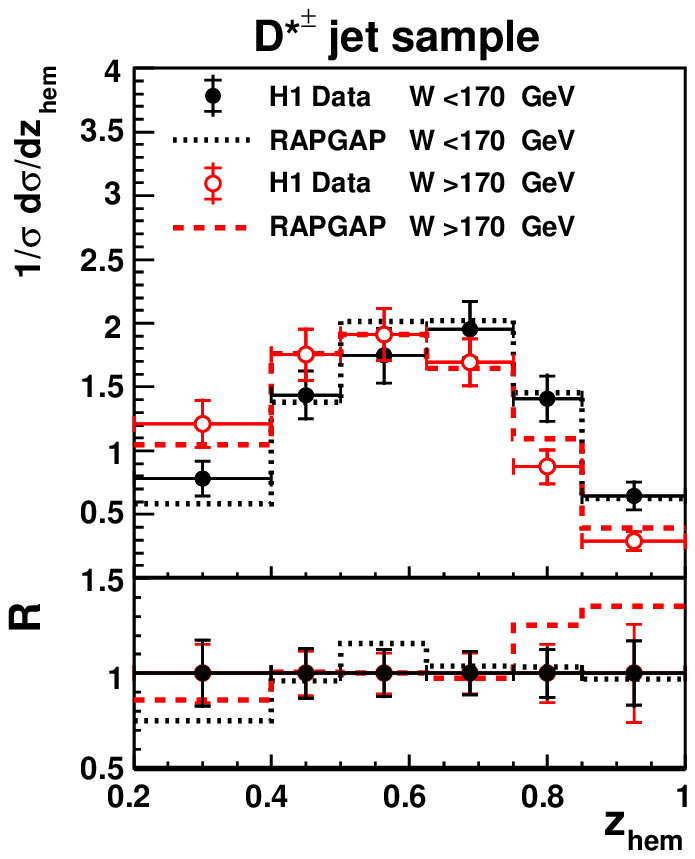,width=0.49\textwidth}
\setlength{\unitlength}{1cm}
\caption{ Normalised $D^{*\pm}$ meson cross sections as a function of 
${\rm z}_{\rm hem}$ and ${\rm z}_{\rm jet}$ for the $D^{*\pm}$ jet sample
in two regions of $W$, for $W < 170$~GeV and $W > 170$~GeV. The measurements are normalised to unity in the displayed range of 
${\rm z}_{\rm jet}$ and ${\rm z}_{\rm hem}$, respectively. In addition,
the MC predictions of RAPGAP are shown using the ALEPH setting for PYTHIA
and the fitted fragmentation parameters $\alpha$ for the Kartvelishvili
parametrisation as given in table~\ref{table:results-parameters-hem-jet} for ${\rm z}_{\rm jet}$ and ${\rm z}_{\rm hem}$. 
The ratios ${\rm R= MC/data}$, as described in the caption of 
figure~\ref{fig:results-defaults},
are shown for both regions of $W$ on top of each other.}
\label{fig:w-dependence} 
\end{figure}

The distributions of ${\rm z}_{\rm hem}$ and ${\rm z}_{\rm jet}$ are also 
measured in two bins of $Q^2$ and $W$. In the case of $Q^2$, the bins are
defined as $2 < Q^{2} < 10$~GeV$^2$ and $10 < Q^{2} < 100$~GeV$^2$. The
accessible range in $W$, determined by the cuts on $Q^2$ and $y$, 
corresponds to $70<W<270$~GeV, and the bins are defined as $70 < W < 170$~GeV
and $170 \le W < 270$~GeV. Correction factors and systematic uncertainties for
these samples are determined in the same way as for the full $D^{*\pm}$
jet sample. The data are compared to the QCD models with the PYTHIA
parameter setting including higher excited charm states and using the Kartvelishvili fragmentation function. For the low and high $Q^{2}$
bins the measured distributions are found to be almost the same and well
described by the QCD models. The distributions of ${\rm z}_{\rm jet}$ 
for the low and high $W$ regions are also similar. A difference is 
observed for the ${\rm z}_{\rm hem}$ distribution, which is softer at
high $W$ as shown in figure~\ref{fig:w-dependence}. RAPGAP and CASCADE
show the same behaviour as a function of $W$ as observed in data. This
behaviour can be understood as being partly due to enhanced gluon
radiation at high $W$ and partly due to the kinematic effect of the
requirement $P_{\rm T}(D^{*\pm}) > 1.5$~GeV. For events at low $W$,
where the charm quark tends to have smaller energy than at high $W$, a
$D^{*\pm}$ meson needs to carry a large fraction of the original quark
momentum in order to pass the $P_{\rm T}$ requirement.
\par
\begin{figure}[hhh]
\center
\epsfig{file=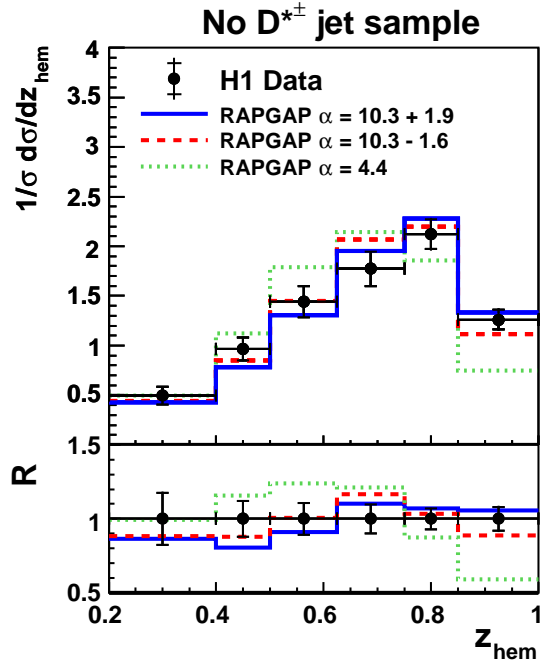,width=0.49\textwidth}
\setlength{\unitlength}{1cm}
\caption{ Normalised $D^{*\pm}$ meson cross sections as a function of 
${\rm z}_{\rm hem}$ for the no  $D^{*\pm}$ jet sample. The measurements
are normalised to unity in the displayed range of ${\rm z}_{\rm hem}$.
The data are compared to MC predictions of RAPGAP with the ALEPH setting for PYTHIA and the Kartvelishvili fragmentation function using the
fragmentation parameter $\alpha$ fitted according to the procedure
described in section~\ref{Section:Results}. The full and dashed lines
indicate a variation of the fragmentation parameter by $\pm 1 \sigma$
around the best fit value of $\alpha$. The dotted line shows the
prediction of RAPGAP with the fragmentation parameter $\alpha = 4.4$
extracted from the $D^{*\pm}$ jet sample. The ratio ${\rm R = MC/data}$
is described in the caption of figure~\ref{fig:results-defaults}.}
\label{fig:results-rapgap-kart-twosamples} 
\end{figure}

The hemisphere observable allows an investigation of charm fragmentation
close to the kinematic threshold, at the limit of applicability of the
concept of fragmentation functions, by selecting events without a 
$D^{*\pm}$ jet with $E_{\rm T}^* > 3$~GeV. As estimated by MC, the mean
centre-of-mass energy squared of the $\gamma^*g$ system $\hat{s}$ 
for this sample is about $36$~GeV$^2$, to be compared with 
about $100$~GeV$^2$ for the $D^{*\pm}$ jet sample. This event sample 
(no $D^{*\pm}$ jet sample) has no overlap with the $D^{*\pm}$ jet sample investigated so far. The normalised cross section as a function of 
${\rm z}_{\rm hem}$ for the no $D^{*\pm}$ jet sample is shown in 
figure~\ref{fig:results-rapgap-kart-twosamples} and listed in table~\ref{table:results-hem-nodsjet}. Predictions of RAPGAP with the three
commonly used fragmentation parameter settings for PYTHIA (see
table~\ref{table:steering}), which provide a reasonable description of the 
$D^{*\pm}$ jet sample (see table~\ref{table:results-parameters-default}),
fail for this sample. Also the prediction using the fragmentation
parameters obtained from the $D^{*\pm}$ jet sample is not able to describe
these data. 
\par
\begin{figure}[hhh]
\center
\epsfig{file=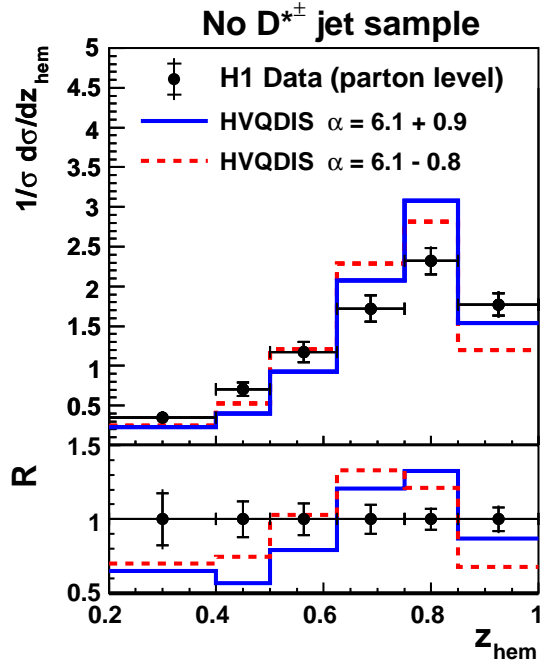,width=0.49\textwidth}
\setlength{\unitlength}{1cm}
\caption{ Normalised $D^{*\pm}$ meson cross sections as a function of 
${\rm z}_{\rm hem}$ for the no  $D^{*\pm}$ jet sample. The data are corrected for hadronisation effects (see section~\ref{Section:QCDModels}). 
The measurements are normalised to unity in the displayed range of 
${\rm z}_{\rm hem}$. The data are compared to NLO predictions of HVQDIS
with the Kartvelishvili fragmentation function using the fragmentation
parameter $\alpha$ fitted according to the procedure described in 
section~\ref{Section:Results}. The full and dashed lines indicate a
variation of the fragmentation parameter by $\pm 1 \sigma$ around the 
best fit value of $\alpha$. The ratio ${\rm R=  MC/data}$ is described
in the caption of figure~\ref{fig:results-defaults}.}
\label{fig:results-hvqdis-kart-nodssample} 
\end{figure}

The fragmentation parameters for RAPGAP, CASCADE and the NLO calculation
are extracted from the no $D^{*\pm}$ jet sample using the same procedure
as for the $D^{*\pm}$ jet sample. The fit results are summarised in
table~\ref{table:results-parameters-nodsjet}.
The predictions of RAPGAP, showing the $\pm 1\sigma$ total uncertainty
around the fitted value of $\alpha$ are also presented in 
figure~\ref{fig:results-rapgap-kart-twosamples}. The fragmentation parameters obtained for RAPGAP and CASCADE are statistically 
compatible. The fragmentation parameters fitted to the no $D^{*\pm}$ jet sample are found to be significantly different from those for
the $D^{*\pm}$ jet sample. They indicate that the fragmentation function
for an optimal description of the sample without a $D^{*\pm}$ jet needs
to be significantly harder than for the  $D^{*\pm}$ jet sample. The NLO
calculation as implemented in HVQDIS fails to describe the no $D^{*\pm}$
jet sample as shown in figure~\ref{fig:results-hvqdis-kart-nodssample}.
\par
Several parameters of the QCD models, for example those influencing
parton showers, have been varied, in trying to describe both samples 
using the same value for the fragmentation function parameter. However,
it was not possible to find MC parameters leading to a consistent
fragmentation function for the two samples. 
Furthermore, the effect of diffractive production of $D^{*\pm}$ mesons
was not able to explain the difference between the fragmentation 
parameters observed for the two samples.
These investigations indicate that QCD models, together with simple parametrisations of the fragmentation functions, are not able to
describe charm fragmentation consistently in the full phase space down
to the kinematic threshold.

\section{Conclusions}

The fragmentation of charm quarks into $D^{*\pm}$ mesons in DIS is studied
using the H1 detector at the HERA collider. The normalised $D^{*\pm}$
meson differential cross sections are measured as a function 
of two observables sensitive to fragmentation, the hemisphere observable ${\rm z}_{\rm hem}$
and the jet observable ${\rm z}_{\rm jet}$, in the visible 
DIS phase space defined by $2 < Q^{2} < 100$~GeV$^{2}$ and $0.05 < y < 0.7$,
and the $D^{*\pm}$ meson phase space defined by $1.5 < P_{\rm T}(D^{*\pm}) <
15$~GeV and 
$| \eta(D^{*\pm}) | < 1.5$. An additional jet with $E_{\rm T}^* > 3$~GeV,
containing the $D^{*\pm}$ meson, is required in the $\gamma^{*}p$
rest-frame in order to provide a hard scale for the events.
\par  
The data are compared with predictions of RAPGAP with three widely used PYTHIA parameter settings and the Peterson and the Bowler parametrisations
for the fragmentation of heavy flavours obtained from $e^{+}e^{-}$
annihilation. They provide a reasonable description of the $ep$ data
presented. 
\par
The normalised differential cross sections are used to fit the parameters
of the Kartvelishvili and Peterson fragmentation functions within the
framework of the QCD models RAPGAP and CASCADE. The fragmentation
parameters extracted using the ${\rm z}_{\rm hem}$ and ${\rm z}_{\rm jet}$ 
observables are in good agreement with each other. Both QCD models lead 
to statistically compatible parameters. The value of the Peterson 
parameter $\varepsilon$ extracted for the PYTHIA parameter setting,
which includes not only $D^{*\pm}$ mesons from direct fragmentation of 
charm quarks but also from the decays of higher excited charm states,
is in agreement with the value of $\varepsilon = 0.04$ tuned by 
ALEPH. 
This result is consistent with the hypothesis of fragmentation 
universality between $ep$ and $e^+e^-$ collisions. 
\par
The QCD models, with the fragmentation parameters fitted to the data, also
provide a good description of the $Q^2$ and $W$ dependence of the 
fragmentation observables.
\par
The data, corrected to the parton level, are also compared to the NLO
calculation as implemented in HVQDIS, with the addition of independent fragmentation of charm quarks to $D^{*\pm}$ mesons. A good fit to the data is obtained when using the fragmentation function by Kartvelishvili 
et al., while using the one of Peterson et al. results in a poor fit. 
\par
Finally, the hemisphere method is used to study the fragmentation of 
charm produced close to the kinematic threshold, by selecting a data 
sample fulfilling the nominal requirements on the DIS and $D^{*\pm}$
meson phase space, but without a $D^{*\pm}$ jet having 
$E_{\rm T}^* > 3$~GeV in the event. The fragmentation parameters
extracted for the QCD models, using this sample of events, are
significantly different from those fitted to the $D^{*\pm}$ jet sample.
Furthermore, the fit for the NLO calculation using the no $D^{*\pm}$ 
jet sample fails. Both observations can be interpreted as an inadequacy
of the QCD models and the NLO calculation to provide a consistent
description of the full phase space down to the kinematic threshold.

\section*{Acknowledgements}

We are grateful to the HERA machine group whose outstanding efforts
have made this experiment possible. We thank the engineers and
technicians for their work in constructing and maintaining the H1
detector, our funding agencies for financial support, the DESY
technical staff for continual assistance and the DESY directorate for
support and for the hospitality which they extend to the non DESY
members of the collaboration.

\newpage
%
\begin{table}[htbp]
  \begin{center}
    \begin{tabular}{|l|r|r|l|}
\hline
PYTHIA     & ALEPH    & Default  & Description \\
parameter  & setting  & setting  &             \\      
\hline      
MSTJ($12$) &   $2$    &   $2$    & baryon model option  \\
MSTJ($46$) &   $0$    &   $3$    & parton shower azimut. corr. \\
MSTJ($51$) &   $0$    &   $0$    & Bose-Einstein correlations off \\
PARJ($1$)  &  $0.108$ &  $0.100$ & P(qq)/P(q)\\
PARJ($2$)  &  $0.286$ &  $0.300$ & P(s)/P(u) \\
PARJ($3$)  &  $0.690$ &  $0.400$ & P(us)/P(ud)/P(s)/P(d)  \\
PARJ($4$)  &  $0.050$ &  $0.050$ & ($1/3$)P(ud\_$1$)/P(ud\_$0$) \\
PARJ($11$) &  $0.553$ &  $0.500$ & P(S=$1$)d,u \\
PARJ($12$) &  $0.470$ &  $0.600$ & P(S=$1$)s   \\
PARJ($13$) &  $0.650$ &  $0.750$ & P(S=$1$)c,b \\
PARJ($14$) &  $0.120$ &  $0.000$ & P(S=$0$,L=$1$,J=$1$) AXIAL  \\
PARJ($15$) &  $0.040$ &  $0.000$ & P(S=$1$,L=$1$,J=$0$) SCALAR \\
PARJ($16$) &  $0.120$ &  $0.000$ & P(S=$1$,L=$1$,J=$1$) AXIAL  \\
PARJ($17$) &  $0.200$ &  $0.000$ & P(S=$1$,L=$1$,J=$2$) TENSOR \\
PARJ($19$) &  $0.550$ &  $1.000$ & extra baryon suppression  \\
PARJ($21$) &  $0.366$ &  $0.360$ & $\sigma_q$ \\
PARJ($25$) &  $1.000$ &  $1.000$ & extra $\eta$ suppression \\
PARJ($26$) &  $0.276$ &  $0.400$ & extra $\eta'$ suppression \\
PARJ($41$) &  $0.400$ &  $0.300$ & Lund symm. fragm.: a \\
PARJ($42$) &  $0.885$ &  $0.580$ & Lund symm. fragm.: b \\
PARJ($54$) & $-0.040$ & $-0.050$ & Peterson fragm.: $-\varepsilon_c$ \\
PARJ($55$) & $-0.002$ & $-0.005$ & Peterson fragm.: $-\varepsilon_b$ \\
PARJ($82$) &  $1.390$ &  $1.000$ & $Q_0$ \\
PARP($72$) &  $0.295$ &  $0.250$ & $\Lambda$ for $\alpha_s$ in time-like
parton \\
           &          &          & showers \\
\hline
    \end{tabular}
\caption{ PYTHIA (version $6.2$) parameter settings: 
ALEPH~\cite{ALEPH-steering} and default. The ALEPH \mbox{setting}
together with the Peterson fragmentation function is used for detector
corrections. \mbox{Detailed} explanation of the parameters can be found 
in~\cite{PYTHIA62}. }
    \label{table:steering}
  \end{center}
\end{table}
%
\renewcommand{\arraystretch}{1.4} 
\begin{table}[htbp]
  \begin{center}
    \begin{tabular}{|l|c|c|c|c|}
\hline
        &  RAPGAP & CASCADE &  HERWIG &  HVQDIS \\ 
\hline
\hline
Proton PDFs   & CTEQ5L~\cite{CTEQ5}  & A0~\cite{CASCADE-gluon} &
CTEQ5L~\cite{CTEQ5}  & CTEQ5F3~\cite{CTEQ5} \\
Photon PDFs   & SaSD-2D~\cite{SaSgam} &  - & SaSG-1D~\cite {SaSgam} &   -	    \\
\hline  
Renorm. scale $\mu_r$    &  $\sqrt{Q^{2}+P_{\rm T}^{*2}}$  &  
$\sqrt{4m_{c}^{2}+ P_{\rm T}^{*2}}$ & 
$\sqrt{\hat{s}}$      &  $\sqrt{Q^{2}+4m_{c}^{2}}$ \\
Fact. scale $\mu_f$   &  $\sqrt{Q^{2}+P_{\rm T}^{*2}}$  & 
$\sqrt{\hat{s}+Q^{*2}_{\rm T}}$     &  
$\sqrt{\hat{s}}$      &  $\sqrt{Q^{2}+4m_{c}^{2}}$  \\
\hline  
Fragmentation model& Lund string  & Lund string & cluster & independent \\
\hline
\end{tabular}
\caption{ Parton density functions (PDFs), fragmentation models and
basic parameters used in the QCD models and the NLO calculation. The 
mass $m_{c}$ of the charm quark is $1.5$~GeV in all cases. The transverse momentum of the charm quark in the $\gamma^* p$ rest-frame is given by 
$P_{\rm T}^{*}$. The invariant mass squared and the transverse momentum 
squared of the $c\bar{c}$ pair are denoted by $\hat{s}$ and 
$Q_{\rm T}^{*2}$, respectively. }
\label{table:models}
\end{center}
\end{table}
%
\begin{table}[htb]
\begin{center}
\begin{tabular}{|l|c|c|c|}
\hline
  & \multicolumn{2}{|c|}
  {$D^{*\pm}$ jet sample}     & No $D^{*\pm}$ jet sample \\
\hline
  & ${\rm z}_{\rm hem}$ error & ${\rm z}_{\rm jet}$ error
  & ${\rm z}_{\rm hem}$ error  \\
\hline
\hline
Statistical uncertainty                & $9.5$\% & $10.9$\%  & $10.9$\% \\
\hline \hline
Source of systematic uncertainty       &         &           &         \\
\hline
Scattered positron energy scale        & $0.8$\% & $0.5$\%   & $0.5$\% \\
Positron scattering angle              & $0.1$\% & $0.1$\%   & $0.1$\% \\
Hadronic energy scale                  & $3.0$\% & $2.5$\%   & $2.1$\% \\
Track reconstruction efficiency        & $0.1$\% & $0.1$\%   & $0.1$\% \\
d$E/$d$x$ measurement                  & $0.1$\% & $0.3$\%   & $0.8$\% \\
$D^{*\pm}$ signal extraction           & $3.0$\% & $3.0$\%   & $2.3$\% \\	
Beauty fraction                        & $1.2$\% & $0.9$\%   & $0.6$\% \\
Migrations into the visible phase space
                                       & $0.1$\% & $0.3$\%   & $0.1$\% \\
\hline
Total systematic uncertainty (hadron level) 
                                       & $4.6$\% & $4.2$\%   & $3.5$\% \\
\hline \hline
Hadronisation effects                  & $3.9$\% & $9.6$\%   & $2.4$\% \\
Total syst. uncertainty (parton level) & $6.3$\% & $11.1$\%  & $4.6$\% \\
\hline
\end{tabular}
\caption{ Experimental and model systematic uncertainties of the
normalised $\rm z$ distributions, \mbox{averaged} over all bins. 
The last two
uncertainties in the table apply only when data are additionally
corrected for hadronisation effects to be compared with HVQDIS. The 
table also provides the statistical errors, averaged over all bins, for
comparison with the systematic uncertainties.}
\label{table:systematics}
\end{center}
\end{table}
\clearpage
%
\renewcommand{\arraystretch}{1.30} 
\begin{table}[htbp]
\begin{center}
{
\scriptsize
\begin{tabular}{|c|c|c|c|ccc|c|}
\hline
\multicolumn{8}{|c|}{\small $D^{*\pm}$ jet sample: ${\rm z}_{\rm hem}$ } \\
\hline      
\multirow{2}{*}{Bin in ${\rm z}_{\rm hem}$} 
 & \multirow{2}{*}{$\frac{1}{\sigma} \frac{{\rm d} 
   \sigma}{{\rm d}{\rm z}_{\rm hem}}$}
 & Statistical & Uncorrelated   
 & \multicolumn{3}{|c|}{Correlated systematic errors}
 & \multirow{2}{*}{Total error} \\
 &   & error  & systematic error & Positron energy & Hadronic scale
 & Beauty &  \\
\hline
$[0.2-0.4[$    & $0.93$ & $0.11$ & $0.03$ & $-0.008$ & $-0.027$ & $-0.020$
  & $0.12$ \\
$[0.4-0.5[$    & $1.53$ & $0.13$ & $0.05$ & $-0.014$ & $-0.037$ & $-0.019$
  & $0.15$ \\
$[0.5-0.625[$  & $1.80$ & $0.15$ & $0.05$ & $+0.002$ & $-0.032$ & $+0.007$
  & $0.16$ \\
$[0.625-0.75[$ & $1.85$ & $0.14$ & $0.06$ & $+0.008$ & $+0.030$ & $+0.019$
  & $0.16$ \\
$[0.75-0.85[$  & $1.27$ & $0.11$ & $0.04$ & $+0.008$ & $+0.056$ & $+0.016$
  & $0.13$ \\ 
$[0.85-1.0]$   & $0.52$ & $0.06$ & $0.02$ & $+0.010$ & $+0.025$ & $+0.007$
  & $0.07$ \\
\hline
    \end{tabular}  
}
\caption{{\rm Normalised $D^{*\pm}$ meson differential cross 
sections as a function of ${\rm z}_{\rm hem}$ for the $D^{*\pm}$ jet
sample, in the visible phase space described in 
section~\ref{Subsection:results_default}. The measurements are normalised
such that their integral over the ${\rm z}_{\rm hem}$ range yields unity.
All errors are considered to be symmetric in each bin. For correlated systematic errors a relative sign is indicated. }}
    \label{table:results-hem-dsjet}
  \end{center}
\end{table}
%
\renewcommand{\arraystretch}{1.30} 
\begin{table}[htbp]
\scriptsize
  \begin{center}
    \begin{tabular}{|c|c|c|c|ccc|c|}
\hline
\multicolumn{8}{|c|}{\small $D^{*\pm}$ jet sample: ${\rm z}_{\rm jet}$ } \\
\hline      
\multirow{2}{*}{Bin in  ${\rm z}_{\rm jet}$} 
 & \multirow{2}{*}{$\frac{1}{\sigma} 
   \frac{{\rm d} \sigma}{{\rm d}{\rm z}_{\rm jet}}$} 
 & Statistical & Uncorrelated   
 & \multicolumn{3}{|c|}{Correlated systematic errors} 
 & \multirow{2}{*}{Total error} \\
 &  & error & systematic error & Positron energy & Hadronic scale
 & Beauty &  \\
\hline  
$[0.3-0.55[ $ & $0.61$ & $0.10$ & $0.02$ & $-0.005$ & $-0.029$ & $-0.016$ & $0.11$ \\
$[0.55-0.7[ $ & $1.76$ & $0.15$ & $0.05$ & $-0.001$ & $-0.026$ & $+0.004$ & $0.16$ \\
$[0.7-0.825[$ & $2.17$ & $0.18$ & $0.07$ & $+0.009$ & $+0.032$ & $+0.017$ & $0.20$ \\
$[0.825-0.9[$ & $1.47$ & $0.18$ & $0.05$ & $-0.011$ & $+0.043$ & $+0.004$ & $0.19$ \\
$[0.9-1.0]  $ & $2.03$ & $0.17$ & $0.06$ & $+0.012$ & $+0.038$ & $+0.011$ & $0.19$ \\
\hline
    \end{tabular}
\caption{{\rm Normalised $D^{*\pm}$ meson differential cross 
sections as a function of ${\rm z}_{\rm jet}$ for the $D^{*\pm}$ jet
sample, in the visible phase space described in 
section~\ref{Subsection:results_default}. The measurements are normalised
such that their integral over the ${\rm z}_{\rm jet}$ range yields unity.
All errors are considered to be symmetric in each bin. For correlated
systematic errors a relative sign is indicated. }}
    \label{table:results-jet}
  \end{center}
\end{table}
%
\renewcommand{\arraystretch}{1.1} 
\begin{table}[htbp]
  \begin{center}
    \begin{tabular}{|c|c|c|c|c|}
       \hline       
\multicolumn{2}{|l|}{ } & \multicolumn{2}{|c|}{ $D^{*\pm}$ jet} & 
No $D^{*\pm}$ jet \\
\multicolumn{2}{|l|}{ } & \multicolumn{2}{|c|}{ sample} & 
sample \\
\hline
\multicolumn{2}{|c|}{RAPGAP using fragmentation by PYTHIA}    
 & Hemisphere & Jet  & Hemisphere \\
Parameter setting & Fragmentation function & ($\chi^2$/n.d.f.) 
 & ($\chi^2$/n.d.f.) & ($\chi^2$/n.d.f.) \\
\hline 
Aleph   &  Peterson $\varepsilon= 0.04$  & $5.1/5$ & $4.0/4$ & $34.2/5$ \\
Default &  Peterson $\varepsilon= 0.05$  & $5.3/5$ & $5.8/4$ & $30.0/5$ \\
Default &  Bowler $a= 0.3, b=0.58$       & $4.8/5$ & $3.3/4$ & $21.1/5$ \\
\hline
    \end{tabular}
\caption{ {\normalsize The PYTHIA parameter settings and fragmentation
functions used for the RAPGAP predictions and the corresponding values
of $\chi^2$/n.d.f. for the $D^{*\pm}$ jet as well as the no $D^{*\pm}$
jet data samples, in the visible phase space described in 
section~\ref{Subsection:results_default}.} }
    \label{table:results-parameters-default}
  \end{center}
\end{table}
\renewcommand{\arraystretch}{1.2} 
\begin{table}[htbp]
  \begin{center}
    \begin{tabular}{|c|c|c|c|c|}
\hline
\multicolumn{5}{|c|}{$D^{*\pm}$ jet sample}  \\  
\hline       
\multirow{2}{*}{Model}    
 & \multicolumn{2}{|c|}{$\alpha \;$ Kartvelishvili } 
 & \multicolumn{2}{|c|}{$\varepsilon \;$ Peterson } \\
 & \multicolumn{2}{|c|}{ {\footnotesize($\chi^2$/n.d.f.)}} 
 & \multicolumn{2}{|c|}{ {\footnotesize($\chi^2$/n.d.f.)}} \\
\hline
 & Hemisphere & Jet & Hemisphere & Jet \\
\hline \hline
\multicolumn{5}{|l|}{PYTHIA default parameter setting:} \\
\hline
\multirow{2}{*}{RAPGAP}    
 & $\alpha=3.3^{+0.4}_{-0.4}$ & $\alpha=3.1^{+0.3}_{-0.3}$      
 & $\varepsilon=0.049^{+0.012}_{-0.010}$
 & $\varepsilon=0.061^{+0.011}_{-0.009}$   \\
 & {\footnotesize $(1.6/4)$ } & {\footnotesize $(2.2/3)$} 
 & {\footnotesize $(5.3/4)$}  & {\footnotesize $(4.2/3)$} \\
\hline
\multirow{2}{*}{CASCADE}  
 & $\alpha=3.5^{+0.5}_{-0.4}$ & $\alpha=3.2^{+0.3}_{-0.3}$      
 & $\varepsilon=0.045^{+0.012}_{-0.009}$   
 & $\varepsilon=0.060^{+0.011}_{-0.009}$   \\ 
 & {\footnotesize $(2.1/4)$}  & {\footnotesize $(3.2/3)$} 
 & {\footnotesize $(5.7/4)$}  & {\footnotesize $(4.7/3)$} \\
\hline \hline
\multicolumn{5}{|l|}{PYTHIA with ALEPH parameter setting:} \\
\hline
\multirow{2}{*}{RAPGAP}    
 & $\alpha=4.4^{+0.6}_{-0.5}$ & $\alpha=4.3^{+0.5}_{-0.4}$      
 & $\varepsilon=0.030^{+0.007}_{-0.006}$     
 & $\varepsilon=0.035^{+0.007}_{-0.006}$     \\
 & {\footnotesize $(3.0/4)$}  & {\footnotesize $(2.8/3)$}  
 & {\footnotesize $(4.0/4)$}  & {\footnotesize $(3.8/3)$} \\
\hline
    \multirow{2}{*}{CASCADE} 
 & $\alpha=4.5^{+0.6}_{-0.6}$ & $\alpha=4.4^{+0.5}_{-0.4}$      
 & $\varepsilon=0.028^{+0.008}_{-0.006}$
 & $\varepsilon=0.034^{+0.007}_{-0.006}$     \\
 & {\footnotesize $(2.4/4)$}  & {\footnotesize $(2.4/3)$}
 & {\footnotesize $(3.3/4)$}  & {\footnotesize $(3.5/3)$} \\
\hline \hline
\multicolumn{5}{|l|}{Fixed-order (NLO) calculation:} \\
\hline
\multirow{2}{*}{HVQDIS}
 & $\alpha=3.3^{+0.4}_{-0.4}$ & $\alpha=3.8^{+0.3}_{-0.3}$
 & $\varepsilon=0.068^{+0.015}_{-0.013}$      
 & $\varepsilon=0.034^{+0.004}_{-0.004}$   \\
 & {\footnotesize $(4.4/4)$}  & {\footnotesize $(4.9/3)$}
 & {\footnotesize $(18.3/4)$} & {\footnotesize $(23.3/3)$} \\
\hline  
    \end{tabular}
\caption{ {\normalsize Fragmentation function parameters extracted for
the QCD models RAPGAP and CASCADE, with the PYTHIA parameter settings as
summarised in table~\ref{table:steering}, and for the NLO QCD program
HVQDIS, using the hemisphere and jet observables measured with the
$D^{*\pm}$ jet sample in the visible phase space described in
section~\ref{Subsection:results_default}. } }
    \label{table:results-parameters-hem-jet}
  \end{center}
\end{table}
%
\renewcommand{\arraystretch}{1.40} 
\begin{table}[htbp]
\scriptsize
  \begin{center}
    \begin{tabular}{|c|c|c|c|ccc|c|}
\hline
\multicolumn{8}{|c|}{\small No $D^{*\pm}$ jet sample: ${\rm z}_{\rm hem}$}
\\
\hline     
\multirow{2}{*}{Bin in ${\rm z}_{\rm hem}$} 
 & \multirow{2}{*}
   {$\frac{1}{\sigma} \frac{{\rm d} \sigma}{{\rm d}{\rm z}_{\rm hem}}$}
 & Statistical & Uncorrelated   
 & \multicolumn{3}{|c|}{Correlated systematic errors}
 & \multirow{2}{*}{Total error} \\
 &   & error  & systematic error & Positron energy & Hadronic scale 
 & Beauty &  \\
\hline
$[0.2-0.4[$    & $0.50$ & $0.09$ & $0.01$ & $+0.003$ & $-0.017$ & $-0.007$ 
 & $0.09$ \\
$[0.4-0.5[$    & $0.97$ & $0.12$ & $0.02$ & $+0.004$ & $-0.023$ & $-0.009$
 & $0.12$ \\
$[0.5-0.625[$  & $1.44$ & $0.16$ & $0.03$ & $-0.006$ & $-0.026$ & $-0.002$
 & $0.16$ \\
$[0.625-0.75[$ & $1.77$ & $0.17$ & $0.04$ & $-0.016$ & $-0.005$ & $+0.005$
 & $0.18$ \\
$[0.75-0.85[$  & $2.13$ & $0.15$ & $0.05$ & $+0.008$ & $+0.037$ & $+0.009$
 & $0.16$ \\ 
$[0.85-1.0]$   & $1.26$ & $0.10$ & $0.03$ & $+0.006$ & $+0.038$ & $+0.006$
 & $0.11$ \\
\hline
    \end{tabular}  
\caption{ {\rm Normalised $D^{*\pm}$ meson differential cross
sections as a function of ${\rm z}_{\rm hem}$ for the no $D^{*\pm}$ jet
sample, in the visible phase space described in
section~\ref{Subsection:results_default}. The measurements are normalised
such that their integral over the ${\rm z}_{\rm hem}$ range yields unity. 
All errors are considered to be symmetric in each bin. For correlated
systematic errors a relative sign is indicated. } }
    \label{table:results-hem-nodsjet}
  \end{center}
\end{table}
%
\renewcommand{\arraystretch}{1.2} 
\begin{table}[htbp]
  \begin{center}
    \begin{tabular}{|c|c|c|}
\hline
\multicolumn{3}{|c|}{ No $D^{*\pm}$ jet sample } \\
\hline
\multirow{1}{*}{ } & \multicolumn{2}{|c|}{Hemisphere observable} \\
\hline      
\multirow{2}{*}{Model} & $\alpha \;$ Kartvelishvili  &
	                     $\varepsilon \;$ Peterson  \\
 & {\footnotesize($\chi^2$/n.d.f.)} & {\footnotesize($\chi^2$/n.d.f.)} \\\hline \hline
\multicolumn{3}{|l|}{PYTHIA default parameter setting:} \\   
\hline
\multirow{2}{*}{RAPGAP} 
 & $\alpha=7.5^{+1.3}_{-1.2}$  & $\varepsilon=0.010^{+0.003}_{-0.003}$ \\
 & {\footnotesize $(5.5/4)$}   & {\footnotesize $(3.9/4)$} \\  
\hline
\multirow{2}{*}{CASCADE}
 & $\alpha=6.9^{+1.1}_{-0.9}$  & $\varepsilon=0.014^{+0.004}_{-0.003}$ \\
 & {\footnotesize $(4.1/4)$}   & {\footnotesize $(2.9/4)$} \\ 
\hline \hline
\multicolumn{3}{|l|}{PYTHIA with ALEPH parameter setting:} \\ 
\hline      
\multirow{2}{*}{RAPGAP}
 & $\alpha=10.3^{+1.9}_{-1.6}$ & $\varepsilon=0.006^{+0.003}_{-0.002}$ \\
 & {\footnotesize $(2.9/4)$}   & {\footnotesize $(1.6/4)$} \\	
\hline
\multirow{2}{*}{CASCADE}       
 & $\alpha=8.4^{+1.3}_{-1.1}$  & $\varepsilon=0.010^{+0.003}_{-0.003}$ \\
 & {\footnotesize $(4.6/4)$}   & {\footnotesize $(4.1/4)$} \\
\hline \hline
\multicolumn{3}{|l|}{Fixed-order (NLO) calculation:}  \\  
\hline \hline
\multirow{2}{*}{HVQDIS}
 & $\alpha=6.1^{+0.9}_{-0.8}$  & $\varepsilon=0.007^{+0.001}_{-0.001}$ \\
 & {\footnotesize $(37.6/4)$}  & {\footnotesize $(38.6/4)$} \\
\hline
    \end{tabular}
\caption{ {\normalsize Fragmentation function parameters extracted for
the QCD models of RAPGAP and CASCADE, with PYTHIA parameter settings as
summarised in table~\ref{table:steering}, and for the NLO QCD program
HVQDIS, using the hemisphere observable measured with the no $D^{*\pm}$
jet sample in the visible phase space described in
section~\ref{Subsection:results_default}.} }
    \label{table:results-parameters-nodsjet}
  \end{center}
\end{table}
%
%
%
%
%
%
%
%
%
\end{document}